\documentstyle[12pt]{article}

\newcommand{\be}{\begin{equation}}
\newcommand{\ee}{\end{equation}}
\newcommand{\dlt}{\delta}
\newcommand{\Dlt}{\Delta}
\newcommand{\ra}{\rightarrow}
\newcommand{\vp}{\varphi}
\newcommand{\bt}{\beta}
\newcommand{\al}{\alpha}
\newcommand{\prt}{\partial}

\newcommand{\om}{\omega}

\newcommand{\lbd}{\lambda}
\newcommand{\gm}{\gamma}
\newcommand{\Gm}{\Gamma}

\newcommand{\ep}{\varepsilon}
\newcommand{\bF}{{\bf F}}
\newcommand{\br}{{\bf r}}
\newcommand{\bL}{{\bf L}}
\newcommand{\bS}{{\bf S}}
\newcommand{\bI}{{\bf I}}
\newcommand{\dgr}{\dagger}

\begin{document}

\begin{center}
{\Large{\bf Principal problems in Bose-Einstein condensation of
dilute gases} \\ [5mm]

V.I. Yukalov} \\ [3mm]

{\it
$^1$Bogolubov Laboratory of Theoretical Physics, \\
Joint Institute for Nuclear Research, Dubna 141980, Russia}

\end{center}

\vskip 2cm

\begin{abstract}

A survey is given of the present state of the art in studying Bose-Einstein
condensation of dilute atomic gases. The bulk of attention is focused on
the principal theoretical problems, though the related experiments are also
mentioned. Both uniform and nonuniform trapped gases are considered. Existing
theoretical contradictions are critically analysed. A correct understanding
of the principal theoretical problems is necessary for gaining a more
penetrating insight into experiments with trapped atoms and for their proper
interpretation.

\end{abstract}

\vskip 2cm

{\parindent=0pt


\vskip 2cm

}

\newpage

\begin{center}
{\bf Contents}
\end{center}

\vskip 2cm

\begin{enumerate}

\item
Introduction

\item
Condensation temperature

\item
Condensate fluctuations

\item
Condensate growth

\item
Low-dimensional condensates

\item
Crossover region

\item
Condensates and superfluids

\item
Collective excitations

\item
Particle correlations

\item
Molecular condensation

\end{enumerate}

\newpage

\section{Introduction}

Bose-Einstein condensation of dilute atomic gases is intensively developing
field of research. There exist several reviews and books on the subject, for
instance [1--5]. However, both experiment and theory in this field develop
so fast that voluminous works are not able to follow the front line of
research. Moreover, the amount of information in the field is so huge that
even detailed reviews do not include all aspects of the problem. The aim
of this survey is to present a brief description of the modern state of the
art related to Bose-Einstein condensation of uniform and trapped dilute
gases. The main concern will be given to the principal problems in
Bose-Einstein condensation. Some of these problems gave rise to controversies.
It is,  therefore, important to present a critical analysis of the basic
theoretical foundations, whose correct understanding is compulsory for the
proper interpretation of experiments with trapped Bose-condensed atoms.

Nowadays, there are several tens of laboratories in different countries,
where Bose-Einstein condensates are routinely produced. Initially, the
condensation of atoms with nonzero spins has been realized, with the
total atomic spin $\bF=\bS+\bL+\bI$ corresponding to a boson, where
$\bS$ is electron spin; $\bL$, electron angular momentum; and with $\bI$
denoting the nuclear spin. The following dilute atomic gases have been
condensed: $^{87}$Rb [6], $^7$Li [7], $^{23}$Na [8], $^{85}$Rb [9], $^{41}$K
[10], $^{133}$Cs [11], H [12], and metastable triplet $^4$He [13,14].
Recently, the spinless atoms of $^{174}$Yb [15] were condensed. There are
indications of producing ultracold condensed molecules $^{85}$Rb$_2$ [16]
formed, by means of Feshbach resonance, in the Bose-Einstein condensate of
$^{85}$Rb. The condensation of molecules $^6$Li$_2$ [17], from the degenerate
Fermi gas of $^6$Li has been reported. More fully the condensation of
molecules will be detailed in the last section below.

In this review, only the systems with Bose-Einstein condensates will be
considered. There exists now an intensive activity of studying degenerate
trapped Fermi gases. These, however, will not be touched here, except the
case when Fermi atoms form Bose molecules that undergo condensation [17].
In the frame of a brief survey, it is impossible to cover all topics related
to cold trapped atoms. So, the emphasis will be given to the principal
theoretical points in Bose-Einstein condensation of dilute atomic gases.

\section{Condensation Temperature}

The influence of atomic interactions on the condensation temperature is a
rather nontrivial problem, because of which much efforts have been devoted
to finding the dependence of the condensation temperature $T_c$ on effective
interaction parameters. The interaction in dilute gases is well characterized
by the Fermi contact potential
$$
\Phi(\br) =  4\pi \; \frac{\hbar^2 a_s}{m}\; \dlt(\br) \; ,
$$
in which $a_s$ is the $s$-wave scattering length and $m$, atomic mass. The
gas is dilute when $|\rho a_s^3|\ll 1$, where $\rho\equiv N/V$ is the number
density. The problem to be solved is how $T_c$ depends on $\rho a_s^2$.

This problem turned out to be complicated even for homogeneous dilute gases,
since Bose-Einstein condensation is a second-order phase transition, which
is governed by long-range correlations. Because of these correlations, simple
mean-field approximations are not always applicable. In an ideal homogeneous
Bose gas, the condensation temperature is
\be
\label{1}
T_0 = \frac{2\pi\hbar^2}{m k_B} \left [ \frac{\rho}{\zeta(3/2)}
\right ]^{2/3} \; ,
\ee
where $\zeta(\cdot)$ is the Riemann zeta function, with $\zeta(3/2)\cong
2.612$. What happens when in the ideal gas the atomic interactions are being
switched on, while the density is kept fixed? The corresponding change of the
condensation temperature can be characterized by the relative difference
\be
\label{2}
\frac{\Dlt T_c}{T_0} \equiv \frac{T_c}{T_0}\; - 1 \; .
\ee
This quantity has been calculated by a variety of methods. Historical
references can be found in [5,18,19]. For repulsive interactions, when the
scattering length is positive and changes from zero to a finite value $a_s$,
the generally accepted functional dependence of the relative shift (2) on the
{\it diluteness parameter} or {\it gas parameter}
$$
\dlt \equiv \rho^{1/3} a_s =  \frac{a_s}{a_0}
$$
is given by the asymptotic expansion
\be
\label{3}
\frac{\Dlt T_c}{T_0} \simeq c_1 \dlt +\left (c_2'\ln\dlt + c_2''
\right )\; \dlt^2 \; ,
\ee
as $\dlt\ra 0$. The first-order coefficient $c_1$ depends on nonperturbative
correlation effects. Its most accurate value was found by numerical Monte
Carlo calculations for three-dimensional lattice O(2) field theory, which
gave $c_1=1.29\pm 0.05$ [20,21] and $c_1=1.32\pm 0.02$ [22,23]. The
coefficient $c_2'$ can be computed exactly using perturbation theory [24],
which yields
$$
c_2' = -\; \frac{64\pi\zeta(1/2)}{3[\zeta(3/2)]^{5/3}} =
19.7518 \; .
$$
The remaining second-order coefficient $c_2''$ depends on nonperturbative
physics and can also be extracted from numerical lattice simulations of
three-dimensional O(2) scalar field theory [23,24], resulting in
$c_2''=75.7\pm 0.4$. Summarizing these results, one has
$$
c_1=1.32\pm 0.02\; , \qquad c_2'=19.7518\; , \qquad
c_2''=75.7\pm 0.4 \; .
$$
There have also been many attempts to calculate the nonperturbative
coefficients $c_1$ and $c_2''$ in different approximations. Thus, within
the frameworks of Ursell operators and of Green functions [18], the
results are $c_1=1.9\; , c_2'=5.589$, and $c_2''=4.9$, which are rather
far from the correct values. Renormalization-group calculations give
$c_1=1.23$ [25] and $c_1=1.15$ [26], but $c_2''$ was not calculated. The
$1/n$-expansion in the $n$-component field theory has also been used,
yielding in the first and second orders, respectively, $c_1=2.33$ [27]
and $c_1=1.71$ [28], but, again, $c_2''$ was not computed. We shall not
enumerate here many other attempts, trying to find the coefficients in
the shift (3), but whose results are not sufficiently reliable (see
discussion in [18,29]).

The most accurate results, obtained by means of an analytical approach,
and being closest to the Monte Carlo numerical simulations, are due to
the usage of the optimized perturbation theory. The main idea of this
theory, whose general formulation was first given in [30] (see further
references in survey [31]), is the introduction of control functions
governing the convergence of the sequence of optimized approximants.
Different variants of the optimized perturbation theory are now widely
used for various problems [31], in particular, in quantum field theory
[32--38]. There are several ways of introducing control functions.

A straightforward method is the introduction of control functions at the
initial stage of a calculational procedure, when they can be incorporated
into the trial Hamiltonian or Lagrangian, or into the trial wave function
or Green function, which the following perturbation theory will be based
on [30]. Control functions are defined at each step of the employed
perturbation theory so that to make convergent the sequence of the
resulting optimized approximants. This method of introducing control
functions into the initial Lagrangian of field theory has been used for
calculating the critical temperature shift in [39--42], where at the
second, third, and fourth orders of the optimized perturbation theory
it was found, respectively, $c_1=3.06,\; 2.45,\; 1.48$ and $c_2''=101.4,
\; 98.2, \; 82.9$. These sequences seemingly converge to the correct
numerical Monte Carlo values. The problem of convergence was specially
studied in [43--46], where the optimized perturbation theory was combined
with $1/n$-expansion. The series are shown to converge exponentially.
The final estimates for the coefficients are $c_1=1.19$ and $c_2''=84.9$.
Thus, the optimized perturbation theory gives the results for both
coefficients $c_1$ and $c_2''$ that exponentially converge to the Monte
Carlo values [39--46].

It is possible to notice that the introduction of control functions into
the initial Lagrangian or Hamiltonian leads to a special scaling of the
calculated quantities. Using this, a way of introducing control functions,
in the frame of the optimized perturbation theory, can be done at the
final stage, when one, first, derives divergent series and then transforms
them, with control functions being incorporated in the process of this
transformation. A particular transformation method has been suggested by
Kleinert [47]. This method was invoked for calculating the coefficient
$c_1$ [48--50]. The calculation, first, starts with deriving loop
expansions for (3+1)-dimensional field theory. The highest expansion,
derived for $c_1$, is the seventh-loop expansion [50]
$$
c_1(g) \simeq a_1 g + a_2 g^2 + a_3 g^3 + a_4 g^4 + a_5 g^5
$$
in powers of an effective coupling parameter, assumed here to be
asymptotically small, $g\ra 0$. The actual $c_1$ corresponds to the limit
$c_1(\infty)$ at $g\ra\infty$. As is clear, the above small-$g$ expansion
has no sense for $g\ra\infty$. To find an effective limit $c_1(\infty)$,
one has to renormalize the above expansion. Such a renormalization can
be  done by introducing control functions by means of the Kleinert
transformation [47] and by defining these control functions from an
optimization procedure. This way also leads to an exponentially fast
convergence of the sequence of optimized approximants [47]. The
coefficients $a_k$ depend on the number of components $n$. For the
two-component field, corresponding to a Bose system, one has [50]
$$
a_1=0.223286\; , \qquad a_2=-0.0661032 \; , \qquad a_3=0.026446\; ,
$$
$$
a_4=-0.0129177\; , \qquad a_5=0.00729073  \qquad (n=2) \; .
$$
Employing the Kleinert method, it is necessary to assume that $c_1(\infty)$
is finite. This limit also essentially depends on the choice of the Wegner
exponent, characterizing the approach to the limit $c_1(\infty)$. Kastening
[49,50] calculated $c_1=c_1(\infty)$ for two different values of the Wegner
exponent. When the latter is defined self-consistently, from the same
expansion, then $c_1=1.376$; but if this exponent is fixed, being chosen
from some additional arguments, then $c_1=1.611$. The average value
$c_1=1.269$ seems to be close to the Monte Carlo results. To check the
method, Kastening [50] has also calculated $c_1(\infty)$ for one-component
and four-component fields, for which Monte Carlo simulations are available
as well [51]: Thus, for $n=1,\; c_1=1.09\pm 0.09$ and for $n=4,\; c_1=1.59
\pm 1.10$. For $n=1$, the expansion coefficients are
$$
a_1=0.334931\; , \qquad a_2=-0.178478 \; , \qquad a_3=0.129786\; ,
$$
$$
a_4=-0.115999\; , \qquad a_5=0.120433  \qquad (n=1) \; .
$$
The seven-loop optimized approximant [50] is $c_1=1.171$ for the
self-consistent choice of the Wegner exponent and $c_1=0.973$, if the latter
is fixed. The average $c_1=1.072$ is close to the Monte Carlo value. In the
case of the four-component field,
$$
a_1=0.167465\; , \qquad a_2=-0.0297465 \; , \qquad a_3=0.00700448\; ,
$$
$$
a_4=-0.00198926\; , \qquad a_5=0.000647007  \qquad (n=4) \; .
$$
The two choices of the Wegner exponent lead [50] to $c_1=1.648$ and
$c_1=1.435$, whose average $c_1=1.542$ is again in agreement with the Monte
Carlo result. The coefficient $c_2''$ was not calculated by this method.

The accuracy of the optimized perturbation theory can be essentially
improved by introducing several control functions. This has recently been
done [52] for calculating the coefficients in the temperature shift (3).
For the coefficient $c_1$, the seven-loop expansion was used, yielding
$c_1=1.3$ and for $c_2''$, the six-loop expansion was employed, giving
$c_2''=73.46$.

The results of all reliable calculations show that the critical
temperature shift (3) is positive and increases with the increasing
diluteness parameter. That is, the temperature of Bose condensation grows
under increasing density or under stronger interaction. Such a behaviour
may look rather strange, though one should not forget that this happens
in the dilute limit, when the diluteness parameter $\dlt\equiv\rho^{1/3}a_s$
is small. When this parameter is large, as for liquid helium, Monte Carlo
simulations show that the critical temperature decreases with growing
density [53]. This is in agreement with the existence of the Bogolubov
depletion, according to which the condensate density $\rho_0$ at zero
temperature is given by the equation
\be
\label{4}
\frac{\rho_0}{\rho} =  1  - \left ( \frac{\dlt}{\dlt_c} \right )^{3/2} \; ,
\ee
where $\dlt\equiv\rho^{1/3} a_s$ is the diluteness parameter and
$$
\dlt_c \equiv \left ( \frac{9\pi}{64} \right )^{1/3} =0.761618 \; .
$$
The Bogolubov depletion formula (4) demonstrates that there exists a
critical value $\dlt_c$ of the diluteness parameter $\dlt$, such that the
condensate density $\rho_0$ becomes zero at the critical temperature $T_c=0$.
The Bogolubov approximation, as is known, is valid for weakly nonideal gases,
when $\dlt\ll 1$. Therefore the critical diluteness parameter $\dlt_c$, when
the complete depletion occurs, may not be very accurately estimated in this
approximation. However Monte Carlo simulations for a wide diapason of $\dlt$
show that the Bogolubov depletion formula is qualitatively correct for
$\dlt\sim 1$ as well, though it is quantitatively accurate up to
$\dlt\approx 0.1$ [54].

In this way, the Bose-Einstein condensation temperature $T_c$ as a function
of the diluteness parameter $\dlt$ behaves as follows. First, when $\dlt$
is asymptotically small, $T_c$ grows with increasing $\dlt$ according to
Eq. (3). But then it turns down, becoming a decreasing function of $\dlt$.
And $T_c$ goes to zero at $\dlt\sim 1$. This type of reentrant behaviour
in the whole range of $\dlt$ has been studied [55] in the Bogolubov
approximation. The maximum $T_c^{max}$ of $T_c$ was found to occur at
$\dlt=\dlt_{max}\approx 0.25\;\dlt_c$, so that
\be
\label{5}
\frac{T_c^{max}}{T_0} \approx 1.1\; , \qquad \dlt_{max} \approx 0.2\; .
\ee
Extracting from [54] the behaviour of $T_c$, as $\dlt$ tends to $\dlt_c$
from below, we get
\be
\label{6}
\frac{T_c}{T_0} \simeq 2.582\sqrt{\dlt_c -\dlt}  \qquad
(\dlt\ra \dlt_c) \; .
\ee
Recall that Eqs. (3) and (6) have to do with a homogeneous
three-dimensional system.

For trapped gases, the situation is quite different. Let the trapping be
realized by the harmonic potential
$$
U(\br) = \frac{m}{2}\; \left ( \om_x^2 x^2 + \om_y^2 y^2 +
\om_z^2 z^2 \right ) \; .
$$
Strictly speaking, there is no sharp phase transition for a finite
number  of atoms $N$. To rigorously define the Bose-Einstein condensation,
one should consider $N\ra\infty$. For the gas of $N$ Bose atoms, trapped
in the harmonic potential, the thermodynamic limit is defined [2,4] as
$$
N\ra\infty\; , \qquad \om_0\ra 0\; , \qquad N\om_0^3\ra {\rm const} \; ,
$$
which follows from the definition of thermodynamic stability [2], and where
the effective frequency is the geometric average
$$
\om_0 \equiv \left ( \om_x\om_y\om_z\right )^{1/3} \; .
$$
In this limit, the Bose-Einstein condensation temperature of an ideal gas
is
\be
\label{7}
T_0 = \frac{\hbar\om_0}{k_B}\left [ \frac{N}{\zeta(3)}
\right ]^{1/3} \; .
\ee
For other types of the trapping potential, the condensation temperature
will be different [2]. Taking into account the final-size corrections,
one has [4] the ideal-gas condensation temperature
\be
\label{8}
T_0' = T_0 -\; \frac{\zeta(2)\hbar\overline\om}{2\zeta(3)k_B} \; ,
\ee
where $\overline\om$ is the arithmetic average
$$
\overline\om \equiv \frac{1}{3}\left ( \om_x + \om_y + \om_z
\right ) \; .
$$

Similarly to the homogeneous case, it is possible to introduce the
diluteness parameter for trapped atoms, whose definition, however, is
different. For the harmonically trapped gas, the diluteness parameter is
given by the ratio
\be
\label{9}
\dlt\equiv a_s/\lbd_0
\ee
of the scattering length $a_s$ to the de Boglie wavelength at the
transition temperature
$$
\lbd_0 \equiv \sqrt{\frac{2\pi\hbar^2}{m k_B T_0}} = \sqrt{2\pi}\;
l_0 \left [ \frac{\zeta(3)}{N}\right ] ^{1/6} \; ,
$$
with $l_0 \equiv \sqrt{\hbar/m\om_0}$ being the oscillator length.
Switching on interatomic interactions [56] changes the condensation
temperature to
\be
\label{10}
\frac{T_c}{T_0} \simeq 1 - \;
\frac{\zeta(2)\overline\om}{2\zeta(3)\om_0}
\left [ \frac{\zeta(3)}{N}\right ]^{1/3} + c_1 \dlt +
\left ( c_2'\ln\dlt + c_2''\right )\dlt^2 \; ,
\ee
where $\dlt$ is asymptotically small. In the case of a uniform Bose gas,
the first-order shift in $T_c$, related to the coefficient $c_1$, is
sensitive to critical fluctuations and so is nonperturbative. In contrast,
the first-order shift for a gas trapped in a harmonic potential is
calculable using perturbation theory [4]. The second-order logarithmic
coefficient $c_2'$ is also calculable in perturbation theory, but the
coefficient $c_2''$ is not [56]. The coefficient $c_2''$ can be calculated
by involving Monte Carlo lattice simulations in three-dimensional O(2)
field theory. The results for these coefficients [56] are
$$
c_1 = -3.426032\; , \qquad c_2'=-\;
\frac{32\pi\zeta(2)}{3\zeta(3)}= - 45.856623\; , \qquad c_2''=-155.
$$
The very first thing that catches the eye, comparing the critical
temperature shifts (3) and (10), is that for a homogeneous gas all
coefficients $c_1,\; c_2'$, and $c_2''$ are positive, while for a trapped
gas all these coefficients are negative. This makes the principal difference
between the uniform and trapped gases. Switching on repulsive interactions
in a uniform gas shifts the condensation temperature up, while in a
trapped  gas this shifts the condensation temperature down. Monte Carlo
investigations [54], of a trapped gas at zero temperature in a wide range
of the diluteness parameter $\dlt$ show that the condensation fraction is
well described by the Bogolubov depletion formula (4) up to $\dlt\sim 0.1$,
but for larger $\dlt\sim 1$, the depletion formula (4) is only
qualitatively correct. It looks like the depletion formula (4) can
be  extended for high $\dlt$ as well, but with the critical diluteness
$\dlt_c$ being different from that given by the Bogolubov approximation.

As is seen, there is no the reentrant effect in trapped gases, in contrast
to uniform gases, where the condensation temperature first increases and
then diminishes as a function of the diluteness parameter. For trapped
gases, $T_c$ is monotonically diminishing function of the gas parameter
$\dlt$.

The formula (10) for the critical temperature of a dilute trapped Bose
gas  is reliable. There were some claims that the linear in $\dlt$ shift
could  be compensated by the finite-size corrections. However, as follows
from  (10),  all shifts have the same negative sign, so there can be
no  compensation between different terms. In addition, formula (10) has
recently been checked for $^{87}$Rb and found to be in good quantitative
agreement with experimental data [57]. The principal difference in the
behaviour of the critical temperature for uniform and trapped gases can
be explained by different properties of fluctuations in these systems.

As will be explained in the following section, the uniform ideal gas is an
unstable system, with anomalously strong particle fluctuations. Switching
on atomic interactions stabilizes the system, making the fluctuations
normal. As a result of such a stabilization, the condensation temperature,
first, increases according to Eq. (3). But further strengthening
interactions leads to the condensate depletion. This is why the
condensation temperature $T_c$ as a function of the diluteness parameter
$\rho^{1/3}a_s$ possesses a specific reentrant behaviour [55].

Contrary to the uniform case, the trapped ideal Bose gas is stable,
having normal fluctuations. Atomic interactions play only the destructive
role, depleting the condensate. Therefore the critical temperature
(10)  is a monotonely decreasing function of the interaction strength
characterized by the parameter $a_s/\lbd_0$ defined in Eq. (9).

\section{Condensate Fluctuations}

Studying fluctuations in statistical systems is important from several
points of view. For instance, fluctuations define the stability of the
system and its way of reaching the state of thermodynamic equilibrium
[58,59].

In the case of Bose-condensed systems, one usually studies the
number-of-particle fluctuations because of their nontrivial properties.
Since in literature, there are controversies related to this problem,
it is necessary to give a careful analysis of the situation.

The number-of-particle fluctuations are characterized by the dispersion
\be
\label{11}
\Dlt^2(\hat N) \equiv\;  <\hat N^2>\; - \; <\hat N>^2
\ee
for the number-of-particle operator
$$
\hat N = \int \psi^\dgr(\br) \psi(\br)\; d\br \; ,
$$
with $\psi(\br)$ being a field operator. The average number of particles
$N$ is a statistical average
$$
N = \; <\hat N>\; = \int \rho(\br)\; d\br \; ,
$$
where $\rho(\br)\equiv<\psi^\dgr(\br)\psi(\br)>$ is the density of particles.
Here we keep in mind an equilibrium state, because of which $\rho(\br)$ does
not depend on time. But $\rho(\br)$ depends on the spatial variable $\br$,
when the system is nonuniform.

The dispersion (11) is directly related to the isothermal compressibility
\be
\label{12}
\kappa_T \equiv -\;
\frac{1}{V} \left ( \frac{\prt V}{\prt P}\right )_{TN} =
\frac{1}{\rho} \left ( \frac{\prt\rho}{\prt P}\right )_{TN}
\ee
by the equality
\be
\label{13}
\kappa_T = \frac{\Dlt^2(\hat N)}{k_B T\rho N} \; ,
\ee
whose derivation can be found, e.g., in [60]. A necessary condition for
a system to be stable is the semi-positiveness and finiteness of the
compressibility, that is, $0\leq\kappa_T<\infty$. If the compressibility
(12) were infinite, this would mean that an infinitesimal fluctuation of
pressure $P$ would lead to an immediate collapse or explosion of the system.
Therefore, in the thermodynamic limit, the dispersion (11) should behave as
\be
\label{14}
\Dlt^2(\hat N) \simeq const\cdot N \qquad (N\ra\infty) \; .
\ee
If $\Dlt^2(\hat N)$ would be proportional to $N$ in a power higher than
one, this would mean that the system is unstable. When the stability
condition (14) is satisfied, the number-of-particle fluctuations are
called normal, but when Eq. (14) is not valid,  so that the compressibility
(13) diverges in the thermodynamic limit, the fluctuations are termed
anomalous.

The dispersion (11) can be presented as
\be
\label{15}
\Dlt^2(\hat N) = \int \left [ R(\br,\br') - \rho(\br)\rho(\br')
\right ] \; d\br\; d\br'
\ee
through the density-density correlation function
$$
R(\br,\br') \equiv \;
<\psi^\dgr(\br)\psi(\br)\psi^\dgr(\br')\psi(\br') > \; .
$$
The latter is related by the equation
$$
R(\br,\br') = \rho(\br)\dlt(\br-\br') + \rho(\br)\rho(\br') g(\br,\br')
$$
with the pair correlation function
$$
g(\br,\br') \equiv
\frac{<\psi^\dgr(\br)\psi^\dgr(\br')\psi(\br')\psi(\br)>}
{\rho(\br)\rho(\br')} \; .
$$
Therefore, another representation of the dispersion (15) is
\be
\label{16}
\Dlt^2(\hat N) = N + \int \rho(\br) \rho(\br') \left [
g(\br,\br') - 1 \right ] \; d\br\; d\br' \; .
\ee
Note that the pair correlation function possesses the property
$g(\br,\br')=g(\br',\br)$. One more form for the dispersion (11) can be
obtained by recollecting the definition of the structural factor
$$
S({\bf k}) \equiv \frac{1}{N} \int \left [ R(\br,\br') -
\rho(\br)\rho(\br')\right ] \; e^{-i{\bf k}\cdot(\br-\br')} \;
d\br\;d\br' \; ,
$$
which can also be written as
$$
S({\bf k}) = 1 + \frac{1}{N}\; \int \rho(\br)\rho(\br')
[ g(\br,\br')-1] \; e^{-i{\bf k}\cdot(\br-\br')} \; d\br\;d\br' \; .
$$
Then, from Eqs. (15) or (16), it follows that
\be
\label{17}
\Dlt^2(\hat N) = N S(0) \; .
\ee
All equations above are valid for any nonuniform system. In the particular
case of a uniform system, when $\rho(\br)=\rho$ and  $g(\br,\br')=
g(\br-\br')$, one has the well known expression
$$
S({\bf k}) = 1 + \rho \int [g(\br) -1 ] \; e^{-i{\bf k}\cdot\br} \; d\br \; .
$$
Then the dispersion (16) reduces to
\be
\label{18}
\Dlt^2(\hat N) = N \left\{ 1 + \rho \int [g(\br) - 1]\;
d\br\right \} \; .
\ee

We see that particle fluctuations are closely related to the
behaviour of the pair correlation function as well as to the structural
factor, which can be measured by means of light or neutron scattering
[61,62]. One more useful relation comes from the expression of the
isothermal compressibility (12) through the isothermal sound velocity
$s$, which reads
$$
\kappa_T = \frac{1}{\rho m s^2} \; , \qquad s^2 \equiv
\frac{1}{m} \left ( \frac{\prt P}{\prt \rho} \right ) _{TN} \; .
$$
Then, for the zero-vector structural factor, one has
$$
S(0) = \rho k_B T \kappa_T = \frac{k_B T}{ms^2} \; .
$$
So that for the dispersion (17), we find
\be
\label{19}
\Dlt^2(\hat N) = \frac{k_B T}{ms^2}\; N \; .
\ee
Note that the same form (19) can be obtained directly from the
Bogolubov theory which provides a good first approximation for dilute
gases. Really, it is easy to show [2] that in the Bogolubov approximation
$$
S(0) = \frac{k_B T}{mc^2} \; , \qquad
c =\sqrt{  \frac{4\pi\hbar^2}{m^2}\; \rho a_s}   \; ,
$$
with $c$ being the sound velocity of elementary excitations. Hence,
we obtain the same form (19), with $s=c$.

Analyzing the relation of the dispersion (18) with the pair
correlation function, we may recall that the general behaviour of
the latter at large $r\equiv|\br|$ is
$$
g(\br) \simeq 1 + \frac{C\; e^{-r/\xi}}{r^{d-2+\eta}}
\qquad (\br\ra\infty) \; ,
$$
where $C$ is a constant, $\xi$ is coherence length, $d$ is space
dimensionality, and $\eta$ is the index of anomalous dimension,
$0\leq\eta\leq 1$. Therefore, for any finite $\xi$, one has
$S(0)\sim{\rm const}$ and $\Dlt^2\hat N\sim N$, that is, the
fluctuations are normal, as it should be for any stable equilibrium
system. Fluctuations can become anomalous only when the coherence
length diverges, $\xi=\infty$, which occurs at the point of a phase
transition. Then $S(0)\sim N^{(2-\eta)/3}$ and
$$
\Dlt^2(\hat N) \sim N^{(5-\eta)/3} \qquad (\xi=\infty) \; .
$$
With the largest value $\eta=1$, we have $\Dlt^2(\hat N)\sim N^{4/3}$.
This means that the fluctuations are anomalous, the compressibility
$\kappa_T\ra\infty$, the sound velocity $s\ra 0$, and the structural
factor $S(0)\ra\infty$. But all this is not a surprise, since the point
of a phase transition is the point of instability. Above as well as
below this point, the coherence length is finite and fluctuations are
normal, which tells that the system is stable. For instance, in the
interacting Bose systems [63,64], the coherence length is $\xi\approx
\hbar/ms$.

Since particle fluctuations are directly related to the behaviour of
the pair correlation function, it is necessary to keep in mind the
general properties of the latter, which follow from those of the
second-order density matrix [65]. The pair correlation function
satisfies the asymptotic boundary condition
$$
g(\br,\br')\ra 1 \qquad (|\br -\br'| \ra \infty)
$$
and the exact normalization condition
$$
\int \frac{\rho(\br)\rho(\br')}{N(N-1)}\; g(\br,\br')\;
d\br d\br' = 1 \; ,
$$
which for the uniform case reduces to
$$
\frac{1}{V} \int g(\br)\; d\br = 1 -\; \frac{1}{N} \; .
$$
In this normalization condition, the thermodynamic limit is assumed.

The fictitious appearance of anomalous fluctuations is often due to
the fact that in the process of calculations the general properties
of the pair correlation function have been occasionally broken. As an
illustrative example, let us adduce the following. From Eq. (18), it
is clear that the major contribution to the integral comes from the
behaviour of $g(\br)$ at large $r\ra\infty$. At large distance, the
interparticle interactions vanish. Then, using the Wick
decoupling for $<\psi^\dgr(\br)\psi^\dgr(0)\psi(0)\psi(\br)$, one
could get
$$
<\psi^\dgr(\br)\psi(\br)><\psi^\dgr(0)\psi(0)>\; + \;
<\psi^\dgr(\br)\psi(0)><\psi^\dgr(0)\psi(\br) > \; .
$$
The density of a Bose-Einstein condensate, $\rho_0$, can be defined
as the limit $|<\psi^\dgr(\br)\psi(0)>|\ra\rho_0$ for $r\ra\infty$.
Consequently, for $g(\br)$ at large $r\ra\infty$, one would get
$1+N_0^2/N^2$, which would result in the anomalous fluctuations
$\Dlt^2(\hat N)\sim N_0^2$. However, the obtained limit for $g(\br)$
is in agreement with the asymptotic boundary condition $g(\br)\ra 1$,
as $r\ra\infty$, only for $N_0\equiv 0$. Hence, actually, there are
no anomalous fluctuations in this case. The contradiction has arisen
because the Wick decomposition cannot be used for correlated systems
with long-range order.

The particle fluctuations below $T_0$ for the {\it ideal} uniform
Bose gas have been studied in a systematic way [66,67]. In the grand
canonical ensemble, the fluctuations of the occupation number
$\hat N_k$ for the operator of the number-of-particles with momentum
${\bf k}$, such that $|{\bf k}|>0$, behave as $\Dlt^2(\hat N_k)\sim
N^{4/3}$. Hence, these fluctuations are anomalous. Even more
anomalous are the fluctuations of the total number of particles
$\Dlt^2(\hat N)=\Dlt^2(\hat N_0)=N_0^2$, which are caused by the
condensate fluctuations. However, as is noted above, such fluctuations
are in agreement with the correct properties of the pair correlation
function only when $N_0=0$. Calculating the dispersion for the number
of particles out of the condensate, $\hat N_{out}\equiv\sum_{k>0}
\hat N_k$, one has [66,67] $\Dlt^2(\hat N_{out})\sim N^{4/3}$. In the
canonical ensemble, when the total number of particles is rigorously
fixed, taking into account the equality $\hat N_0=N-\hat N_{out}$,
one gets $\Dlt^2(\hat N_0)=\Dlt^2(\hat N_{out})$, which is given by
\be
\label{20}
\Dlt^2(\hat N_0) = 4.372 \left ( \frac{T}{T_0}\right )^2 \;
N^{4/3} \; ,
\ee
where $T_0$ is the condensation temperature (1) for the uniform ideal
Bose gas. Formula (20) was first derived by Hauge [68]. At all finite
temperatures, the condensate fluctuations are anomalous.

Thus, the ideal uniform Bose gas is a rather pathological object, with
anomalous fluctuations. It is easy to find its pressure [69], which for
$T<T_0$ is

\be
\label{21}
P_0 = \zeta(5/2)\; \frac{k_B T}{\lbd_T^3} \; , \qquad
\lbd_T \equiv \sqrt{\frac{2\pi\hbar^2}{m k_B T} } \; .
\ee
As is evident, $\prt P_0/\prt\rho=0$, so that the compressibility (12)
diverges, $\kappa_T\ra\infty$, which is in agreement with the anomalous
behaviour of the dispersion $\Dlt^2(\hat N)$. Therefore, the ideal Bose
gas is unstable below $T_0$, though it is stable above this temperature.
For a stable equilibrium system, the compressibility must be finite,
fluctuations normal, and also, the observable quantities, calculated
in different statistical ensembles, must coincide in the thermodynamic
limit. Hence, the ideal uniform Bose gas cannot exist as a stable
equilibrium system below the temperature of condensation.

The theory of weakly nonideal uniform Bose gas was developed by
Bogolubov [70--72]. The dispersion $\Dlt^2(\hat N)$, which he calculated
[71], under the conditions $T<T_0$ and $N-N_0\ll N$, takes the form
\be
\label{22}
\Dlt^2(\hat N) = \frac{k_B T}{mc^2}\; N \; ,
\ee
where $c=\sqrt{4\pi\hbar^2\rho a_s/m^2}$ is the Bogolubov sound velocity.
This is in agreement with the general expression (19). The Bogolubov
theory [70--72] is based on three points. First, the condensate is
described by introducing the condensate field $\psi_0$ by means of the
so-called Bogolubov prescription $\psi=\psi_0+\tilde\psi$ with $\psi_0$
treated as a nonoperator term and $\tilde\psi=\frac{1}{\sqrt{V}}
\sum_{k\neq 0} a_ke^{i{\bf k}\cdot\br}$ being the field operator of
noncondensed atoms. Then, owing to the assumption $N-N_0\ll N$, all
terms of the Hamiltonian containing the factors of $\tilde\psi$, or
$a_k$, of orders higher than two, are omitted. And finally, the
remaining approximate Hamiltonian is diagonalized by means of the
Bogolubov canonical transformation. Since only the terms of up to the
second order in $\tilde\psi$, or $a_k$, are retained in this
approximation, the same has to be done in calculating all observable
quantities. Considering the expressions that are higher than quadratic
in $\tilde\psi$ would go beyond the accuracy of the Bogolubov
approximation and so would be not self-consistent. This was taken into
account by Bogolubov [71] when calculating the dispersion (22) using
the relation (18).

The weakly nonideal Bose gas possesses the pressure [69]
\be
\label{23}
P = P_0 + \frac{2\pi\hbar^2 a_s}{m} \left [ \rho^2 +
\frac{\zeta^2(3/2)}{\lbd_T^6}\right ] \; ,
\ee
where $P_0$ is the pressure (21) of the ideal gas. From here, the
compressibility (12) is $\kappa_T=1/\rho m c$. Therefore, due to Eq.
(13), the particle dispersion coincides with the Bogolubov form (22).

In this way, including atomic interactions makes the fluctuations normal,
thus, stabilizing the system. Returning to the ideal gas would mean
setting $a_s\ra 0$, because of which $c\ra 0$ and $\kappa_T\ra\infty$,
so that the system would become unstable. Equation (22), as is clear,
corresponds to the grand canonical ensemble. In canonical ensemble, the
fluctuations of the total number of particles are not defined, but one
can find the dispersion $\Dlt^2(\hat N_0)$, for the number of condensed
particles, assumed to be equal to $\Dlt^2(\hat N_{out})$. For a weakly
excited interacting gas, condensate fluctuations were found to be normal,
that is, $\Dlt^2(\hat N_0)\sim N$, both in the canonical as well as in
the microcanonical ensembles [73].

In the {\it trapped} ideal Bose gas, particle fluctuations are normal,
contrary to those in the uniform case. Politzer [74] calculated
$\Dlt^2(\hat N_0)$ in both canonical and grand canonical ensembles for
atoms trapped inside a harmonically confining potential, the difference
between the ensembles vanishing in the thermodynamic limit. The Politzer
result is
\be
\label{24}
\Dlt^2(\hat N_0) = \frac{\pi^2}{6\zeta(3)}\left ( \frac{T}{T_0}
\right )^3 \; N \; ,
\ee
where $T_0$ is the condensation temperature (7) for the harmonically
confined ideal gas. Microcanonical ensemble also displays normal
condensate fluctuations [75]. Comparing Eq. (24) with Eq. (20), we see
that the anomalous fluctuations of the ideal uniform gas are suppressed
by the confining  potential.

The influence of interactions in the dilute trapped gas on the size of
fluctuations can be understood by remembering that these interactions
diminish the condensation temperature according to Eq. (10). Therefore,
the dispersion (24) can be transformed to
\be
\label{25}
\Dlt^2(\hat N_0) \approx \frac{\pi^2 N}{6\zeta(3)} \left (
\frac{T}{T_0}\right )^3 \left ( 1 + 10.278\; \rho^{1/3} a_s
\right ) \; .
\ee
Atomic interactions increase the condensate fluctuations in the
trapped gas, leaving the fluctuations normal.

Considering fluctuations in interacting systems, in the frame of the
Bogolubov theory [70--72], it is necessary, as is emphasized above, to
omit in the course of calculations, all terms containing the products
of more than two operators $\tilde\psi$, or $a_k$, corresponding to
noncondensed particles. Retaining such higher-order terms would remove
the consideration outside the region of applicability of the Bogolubov
approach. For example, the dispersion (22), derived in the frame of
this approximation, is normal. However, when calculating the dispersion
$\Dlt^2(\hat N_{out})$, one deals with the fourth-order product of the
operators $a_k$, corresponding to the particles excited out of the
condensate. The direct calculation of such products, with the standard
replacement of the sums over momenta by integration in momentum space,
yields infrared divergencies for both uniform as well as trapped gases.
These actual divergencies can be formally avoided by considering the
discretized sums over the low-energy phonon modes [76], which, anyway,
results in anomalous fluctuations $\Dlt^2(\hat N_{out})\sim N^{4/3}$
for all interacting systems, trapped or untrapped. This anomalous
behaviour arises from the occurrence of infrared divergencies due to
the term containing the fourth-order product of operators $a_k$ of
noncondensed particles [76--78].

It is, by the way, easy to get the $N^{4/3}$ behaviour of
$\Dlt^2(\hat N_{out})$ without discretizing the low-energy phonon
modes. Really, the infrared divergence arising from the improper
usage of the fourth-order operator term, which should be omitted in
the Bogolubov theory, is caused by the integral $N\int dp/p^2$. For
the lower limit of the latter, we could take $p_{min}=\hbar/R$, with
the system radius $R=(3V/4\pi)^{1/3}$. Ten we would immediately get
$\Dlt^2(\hat N_{out})\sim N^{4/3}$.

The operator of the total number of particles is the sum $\hat N=
\hat N_0 +\hat N_{out}$. Therefore
$$
\Dlt^2(\hat N) = \Dlt^2(\hat N_0) + \Dlt^2(\hat N_{out}) +
2\; <\hat N_0 \hat N_{out}>\; - 2N_0 N_{out} \; .
$$
Treating the condensate field as a nonoperator quantity, one has
$\hat N_0=N_0\equiv<\hat N_0>$. Then, in the grand canonical ensemble,
$\Dlt^2(\hat N)=\Dlt^2(\hat N_{out})$. Hence, the anomalous behaviour
of the dispersion $\Dlt^2(\hat N_{out})$ implies the same anomalous
behaviour of $\Dlt^2(\hat N)$. The latter would mean that, in the
thermodynamic limit, the sound velocity becomes zero, compressibility
diverges, $\kappa_T\ra\infty$, and the structural factor
$S(0)\ra\infty$ also becomes infinite everywhere below the transition
temperature. Such anomalies have never been observed neither for
liquid helium nor for trapped gases. Thus, the excitation spectrum
of the trapped Bose-Einstein condensate of $^{87}$Rb atoms has been
thoroughly investigated, including its long-wavelength limit [79].
The excitation spectrum, sound velocity, and the structural factor
were found to be in perfect agreement with the Bogolubov theory
predictions. And no anomalies of these quantities were observed.
In this way, we have to conclude that the anomalous fluctuations
in {\it interacting} Bose systems, with $\Dlt^2(\hat N_{out})\sim
N^{4/3}$, are just artifacts caused by a not self-consistent
calculational procedure.

The problem of the fictitious appearance of anomalous fluctuations
is rather general and occurs for any system with continuous symmetry,
when calculations are not self-consistent. Let us consider a system
with a Hamiltonian containing the vector operators $\bS$, whose rotation
leaves the Hamiltonian unchanged. For concreteness, we may keep in mind
the spin operators. The order parameter is ${\bf M}\equiv<\bS>$. An
external field ${\bf h}$ acts on the operator $\bS$. The influence of
the external field on the order parameter is characterized by the tensor
of susceptibility $\chi_{\al\bt}\equiv\prt M_\al/\prt h_\bt$, whose
general form can be presented as
$$
\chi_{\al\bt} = n_\al n_\bt \chi_{||} + (\dlt_{\al\bt} -
n_\al n_\bt ) \chi_\perp \; ,
$$
where ${\bf n}\equiv{\bf h}/h$ and $h\equiv|{\bf h}|$. Here $\chi_{||}$
is the longitudinal susceptibility, while $\chi_\perp$ is the transverse
susceptibility. Above the critical temperature $T_c$, the system is in
symmetric phase, so that there are no distinct directions of the order
parameter, which implies that $\chi_{||}=\chi_\perp$ for $T>T_c$.

But below $T_c$, the longitudinal and transverse susceptibilities are,
in general, different [80]. Usually, the transverse susceptibility,
in a model with continuous symmetry, is larger than the longitudinal
susceptibility, which results in the dominance of directional over size
fluctuations in destroying order [81]. It may even happen that the
transverse susceptibility diverges as $\chi_\perp\sim h^{-1}$, when
$h\ra 0$. This occurs because, under continuous symmetry, all directions
of the vector order parameter are equivalent, and an infinitesimally
small transverse field may turn the order parameter by a finite angle.
Infinite susceptibility means that the considered model is unstable.
Such an instability can be easily removed by adding to the model
a finite additional field breaking the continuous symmetry. Actually,
any real system always has such breaking-symmetry field. For example,
the isotropic Heisenberg model is well known to be a cartoon of real
magnetic systems, always possessing anisotropic terms due to spin-spin
dipole interactions, spin-orbital interactions, indirect exchange,
crystalline fields of different nature, and demagnetization anisotropic
factors caused by the shape of the sample [82--84]. In the presence of
a finite, though may be weak, field $h$, the transverse susceptibility
$\chi_\perp\sim h^{-1}$ is finite. The related transverse correlation
function $C_\perp=<\dlt S_\perp\dlt S_\perp>$, where $\dlt S_\perp\equiv
S_\perp-<S_\perp>$, has the spatial dependence
$$
C_\perp(r) \sim \frac{\exp(-r/\xi)}{r^{d-2+\eta}} \qquad
(r\ra\infty) \; ,
$$
with the correlation length $\xi\sim h^{-1}$. The corresponding Fourier
transform reads $C_\perp(q)\sim(q^2+\xi^{-2})^{-1}$, as $q\ra 0$. At
zero field $h\equiv 0$, the form $C_\perp(q)\sim 1/q^2$ is known as the
Goldstone theorem, whose rigorous proof for Bose and Fermi systems was
given by Bogolubov [72].

For the longitudinal correlation function $C_{||}=<\dlt S_{||}\dlt
S_{||}>$, with $\dlt S_{||}\equiv S_{||}-<S_{||}>$, one may write [80]
an approximate form
$C_{||}\sim<(\dlt S_\perp)^2(\dlt S_\perp)^2>$. Treating the latter in
the hydrodynamic approximation, one gets [80] the Wick-type decoupling
$$
<(\dlt S_\perp)^2 (\dlt S_\perp)^2 > \; \approx
2< \dlt S_\perp \dlt S_\perp>^2 \; .
$$
This gives the relation $C_{||}(r)\sim C_\perp^2(r)$, from which
it follows that, at zero field $h\equiv 0$, one has $C_{||}(r)\sim
1/r^{2(d-2)}$ and $C_{||}(q)\sim 1/q^{4-d}$. Recall that the index
$\eta=0$ in the hydrodynamic approximation. This tells that the
longitudinal susceptibility $\chi_{||}\sim h^{-1/2}$ also diverges
at zero field [80]. The character of this divergence follows from the
relation $\chi_{||}\sim \int C_{||}(r)d\br\sim N^{(d-2)/3}$, valid for
$2<d<4$. Then the dispersion of the spin operator $\bS$, for which
$\Dlt^2(\bS)\sim N\chi_{||}$, behaves as $\Dlt^2(\bS)\sim N^{(d+1)/3}$.
In the three-dimensional space, one gets $\Dlt^2(\bS)\sim N^{4/3}$ for
all $T<T_c$, that is, the same kind of anomalous fluctuations as for
Bose-Einstein condensates.

If such anomalous fluctuations would really exist, this would mean
that there are no stable ordered phases for the systems with broken
continuous symmetry. This kind of conclusion would certainly be
incorrect. It is well known that stable ordered phases, such as
magnetic phases, do exist [82--85]. The longitudinal susceptibility
for these phases diverges solely at the critical point $T_c$, where
the correlation length behaves as $\xi\sim a|1-T/T_c|^{-\nu}$, with a
critical exponent $\nu$, and $a$ being the mean interparticle distance.
Far below $T_c$, the correlation length at zero field is always finite,
the correlation function exponentially decays, and the longitudinal
susceptibility is also finite, hence, the order-parameter fluctuations
are normal.

The origin of the fictitious anomalous fluctuations in the systems
with continuous symmetry is the same as that occurring in the case of
the Bose-Einstein condensate. The hydrodynamic approximation consists
in leaving in the Hamiltonian only the terms up to second order in the
operators $\dlt S_\perp$. For spin systems, $\dlt S_\perp$ plays the
role of the magnon operators. To calculate $\Dlt^2(\bS)$, one needs
to consider the product of four operators $\dlt S_\perp$, which one
treats by invoking the Wick decoupling and using the hydrodynamic
approximation. However, the four-order operator products cannot be
correctly treated in the second-order hydrodynamic approximation. This
is the same as to try to employ a linear approximation for describing
strongly nonlinear effects. Because of this inconsistence, there arise
infrared divergencies leading to the appearance of fictitious anomalous
fluctuations.

The same inconsistence appears if one tries to calculate the
four-order terms in the particle dispersion of quantum liquids, such
as helium, by using the hydrodynamic approach, which is a second-order
approximation. No anomalous fluctuations arise, when calculations are
self-consistent, as is demonstrated by Bogolubov [71] when considering
the fluctuations in weakly interacting gases with Bose-Einstein
condensate.

Let us now summarize the behaviour of fluctuations in Bose-condensed
systems. The ideal uniform Bose gas, below the condensation temperature,
is an unstable system with anomalous condensate fluctuations described
by the dispersion (20). Switching on atomic interactions stabilizes
the system making fluctuations normal according to Eq. (22).

The trapped ideal Bose gas is stable, with normal fluctuations, as
is described by the dispersion (24) for a harmonic confining potential.
Switching on interactions leaves the fluctuations normal, but increases
them, as is illustrated by Eq. (25). Different properties of ideal
uniform and trapped gases explain why the condensation temperature
in a uniform gas displays the reentrant behaviour as s function of
the scattering length $a_s$, while the condensation temperature in
a trapped gas is a monotonely decreasing function of $a_s$.

\section{Condensate Growth}

Formation of the Bose-Einstein condensed phase from a strongly
nonequilibrium, completely disordered, initial state, ultimately
leading to long-range ordering, is one of the most interesting and
fundamental dynamic problems of the physics of multiparticle systems.
The attempts of describing this process have encountered controversies.
First, Levich and Yakhot [86], using the semiclassical kinetic
equations, concluded that the process of condensation requires
infinite time. To make this time finite, it is necessary to assume
the existence on the nuclei of the condensed phase in the beginning
of the cooling process. Svistunov [87] and Kagan et al. [88--90]
also followed a semiclassical approach based on the nonlinear
Schr\"odinger equation. Their analysis of the kinetics starts with
the statement that the weakly interacting Bose gas possesses large,
relative to unity, occupation numbers and, thus, is essentially
a classical field phenomenon. According to their picture [87--90],
the condensate formation goes through three stages. At the first
stage, substantial suppression of nonequilibrium fluctuations of
the density occurs. Then a kind of short-or medium-range order
appears, when the quasicondensate is formed, such that the density
fluctuations are suppressed, but phase fluctuations are essential.
And the true condensate arises only at the last, third, stage, when
the long-range order develops, with all fluctuations of density as
well as phase being relaxed. Stoof [91,92] stressed the importance
of quantum fluctuations at the initial stage.

Bose-Einstein condensate is a coherent system. The development
of coherence in a nonequilibrium Bose gas is somewhat analogous to
the emergence of coherence in radiating laser-type systems, where
the coherent stage is preceded by the quantum stage, with quantum
fluctuations triggering the appearance of the coherent component
[93,94].

In order to better understand how the condensate is being formed,
it is important to recollect the main characteristic length and
time scales  that are common for all many-particle systems [95--97].

One evident characteristic length is the interaction radius, which
can be well represented by a scattering length $a_s$. This defines
the effective particle velocity $v\sim\hbar/m a_s$, with $m$ being
particle mass. The {\it interaction time}
\be
\label{26}
\tau_{int} = \frac{a_s}{v} \sim \frac{m a_s^2}{\hbar}
\ee
describes the time during which two particles strongly interact with
each other in the course of mutual scattering. The related interval
\be
\label{27}
0 < t < \tau_{int} \qquad (dynamic\; stage)
\ee
is called the dynamic stage, when particles move more or less
independently from each other, their motion being described by
dynamical equations.

The mean free path $\lbd=1/\rho a_s^2$, where $\rho\sim a^{-3}$, with
$a$ being the mean interatomic distance, is the characteristic length
between two scattering events. Typical relations are
$$
\frac{\lbd}{a_s} \sim \frac{1}{\rho a_s^3} \; , \qquad
\frac{\lbd}{a} \sim \left ( \frac{a}{a_s}\right )^2 \; .
$$
The correlation time or {\it local-equilibrium time} is
\be
\label{28}
\tau_{loc} = \frac{\lbd}{v} \sim \frac{m}{\hbar\rho a_s} \; .
\ee
It is connected with the interaction time (26) as
$$
\frac{\tau_{int}}{\tau_{loc}} \sim \rho a_s^3 \; .
$$
The local-equilibrium time (28) can also be expressed through other
characteristic quantities. The healing or coherence length
$\xi=\hbar/m s$ is defined through the sound velocity
$s\sim\hbar\sqrt{\rho a_s}/m$. The latter is related to the particle
velocity $v$ as $s/v\sim\sqrt{\rho a_s^3}$. For the coherence length,
one has
$$
\frac{\xi}{a_s} \sim \frac{1}{\sqrt{\rho a_s^3}} \; , \qquad
\frac{\xi}{a} \sim \sqrt{\frac{a}{a_s}} \; , \qquad
\frac{\xi}{\lbd} \sim \sqrt{\rho a_s^3} \; .
$$
The typical energy scale is given by $\ep\sim\hbar^2\rho a_s/m$.
Therefore, for the local-equilibrium time (28) we get
$$
\tau_{loc} = \frac{\lbd}{v} \sim \frac{\xi}{s} \sim
\frac{\hbar}{\ep} \; .
$$
The temporal interval
\be
\label{29}
\tau_{int} < t < \tau_{loc} \qquad (kinetic \; stage)
\ee
is the kinetic stage, during which particle motion can be described
by kinetic equations. At this stage, because of mutual interactions,
interparticle correlations begin arising.

The existence of the next stage was advanced in [98] (see also [95--97]).
Then the overall system consists of parts inside each of which there
is some kind of ordering, while different parts are not correlated with
each other. The typical size of each part, $l_f$, is mesoscopic, being
in between the mean particle distance $a$ and the linear size of the
whole system, $L$, so that
\be
\label{30}
a \ll l_f \ll L \; .
\ee
Since these mesoscopic subregions are not correlated with each other,
but distributed randomly in space, they are termed heterophase
fluctuations. Examples of the latter are magnetic clusters inside
a paramagnetic matrix, disordered regions inside a crystal, crystalline
formations inside a liquid, superconducting droplets inside a normal
metal, subregions with differing local densities, and so on [95]. In
particular, these could be separated spatial regions, inside each of
which there is a coherent Bose-Einstein condensate, but with no coherence
between the different spatial regions. Denoting the average velocity of
motion for each heterophase nucleus as $v_f$, we may introduce the {\it
heterophase fluctuation time}
\be
\label{31}
\tau_f = \frac{l_f}{v_f} \; ,
\ee
which is a characteristic lifetime of a local heterophase fluctuation.
During the temporal interval
\be
\label{32}
\tau_{loc} \ll t \ll \tau_f \qquad (heterophase \; stage) \; ,
\ee
defining the heterophase stage, the system is spatially nonuniform,
consisting of subregions with different order parameters [95]. The
actual value of $\tau_f$ depends on the particular system considered,
but it is always larger than the local-equilibrium time $\tau_{loc}$,
since to organize a kind of order across the distance $l_f\gg a$
requires a time $\tau_f\gg\tau_{loc}$.

The final temporal interval is the hydrodynamic stage lasting the time
defined by the inequalities
\be
\label{33}
\tau_f \ll t \ll \tau_{exp} \qquad (hydrodynamic \; stage)\; ,
\ee
where $\tau_{exp}$ is the experiment time or the observation time.
In many cases, one may set $\tau_{exp}\ra\infty$. At this final stage,
nonequilibrium systems can be described by hydrodynamic equations.

For usual condense matter, such as solid or liquid, one has
$\rho a_s^3\sim 1$, so that $s\sim v$ and
$a_s\sim a\sim\lbd\sim\xi\ll l_f$. Then the kinetic stage (29) is
practically absent, and only three stages are left, dynamic,
heterophase, and hydrodynamic.

For dilute gases, when $\rho a_s^3\ll 1$, one has $s\ll v$, so
that $a_s\ll a\ll \xi \ll \lbd$. Then all four temporal stages are
present, since $\tau_{int}\ll\tau_{loc}\ll\tau_f$. To estimate the
characteristic scales for dilute gases, let us take the values typical
of experiments with the trapped atoms of $^{87}$Rb and $^{23}$Na,
for which $m\sim 10^{-22}$ g, $a_s\sim 5\times 10^{-7}$ cm, and
$\rho\sim(10^{11}-10^{15})$ sm$^{-3}$. Then the parameter
$\rho a_s^3\sim 10^{-8}-10^{-4}$ is really small. Atomic velocity
is $v\sim 10$ cm/s and sound velocity is $s\sim(10^{-3}- 10^{-1})$ cm/s.
Mean interatomic distance is $a\sim(10^{-5}- 10^{-4})$ cm. Coherence
length is $\xi\sim(10^{-4}-10^{-2})$ cm. Mean free path is $\lbd\sim
(10^{-2}-10^2)$ cm. The interaction time (26) is $\tau_{int}\sim 10^{-8}$
s, that is, the dynamic stage (27) is very short. The local-equilibrium
time (28) is $\tau_{loc}\sim(10^{-4}-1)$ s. Hence, the kinetic stage (29)
is much longer than the dynamic stage (27), because of which the latter
can be neglected in practical description.

In the frame of the defined time scales, one may consider the formation
of the Bose-Einstein condensate from a nonequilibrium disordered state.
Then, according to [87--90], we have the following picture. The dynamic
stage (27), being rather short, is usually omitted from the consideration.
During the kinetic stage (29), substantial suppression of nonequilibrium
fluctuations of the density occurs. After the local-equilibrium time
$\tau_{loc}$, mid-range order develops in regions of size $l_f\geq\xi$.
Different spatial regions are not correlated with each other. Such
a situation is typical of heterophase fluctuations. At the heterophase
stage (32), density fluctuations are suppressed, but phase fluctuations
are essential. The authors [87--90] call such a system {\it
quasicondensate}. After the fluctuation time $\tau_f$, coherence
penetrates through the whole system, true long-range order develops,
so that the Bose-Einstein condensate is finally formed. At the last
hydrodynamic stage (33), the Bose-condensed gas relaxes to equilibrium.

The formation process of a Bose-Einstein condensate in a trap was
described by Gardiner et al. [99,100] using a master equation based
on quantum kinetic theory. The description assumed the existence of
one condensate mode in interaction with an equilibrium thermalized
bath of noncondensate atoms at fixed temperature $T$ and chemical
potential $\mu$. The number of condensed atoms $N_0(t)$ satisfies
the rate equation
\be
\label{34}
\frac{dN_0}{dt} = 2W(N_0) \left \{ 1 + \left [ 1  - exp\left (
\frac{\Dlt\mu}{k_B T}\right ) \right ] \; N_0 \right \} \; ,
\ee
in which $\Dlt\mu=\mu_c(N_0)-\mu$ is the difference between the
condensate chemical potential $\mu_c(N_0)$ and the bath chemical
potential $\mu$, while $W(N_0)$ is a transition probability from the
bath to the condensate. A collision between a pair of atoms initially
in the bath of atomic vapor results in one of the atoms going to the
condensate, because of which $W(0)$ is not zero. Thus, with the initial
condition $N_0(0)=0$, at the initial stage, Eq. (34) gives
$N_0(t)\simeq 2W(0) t$.

Growth of a Bose-Einstein condensate from thermal vapor was
experimentally studied [101] for atoms of $^{87}$Rb. It was found
that there exists a latency time of $(10^{-4}-10^{-3})$ s at which
a condensate is first detected. According to the time scales described
above, the latency time should correspond to the local-equilibrium time
$\tau_{loc}$, since before this time, at the kinetic stage, there cannot
be yet any condensate, which appears after $\tau_{loc}$. Then the slow
linear growth of the condensate was observed [101] till the time of
$5\times 10^{-3}$ s. The latter should correspond to the heterophase
fluctuation time $\tau_f$. The interval $\tau_{loc}<t<\tau_f$ is the
heterophase quasicondensate stage. The numerical calculations based on
the rate equation (34) do not reproduce the measured condensate growth
data accurately in the quasicondensate regime. This equation describes
better the condensate growth at the last hydrodynamic stage, when
$t\gg\tau_f$. Then there is exponential growth till the condensate
reaches its asymptotic equilibrium value at about $10^{-2}$ s. At this
last stage, instead of Eq. (34), one can use the simple form
\be
\label{35}
N_0(t) = N_0(t_0)\left\{ 1 -\exp\left [ -\Gm(t-t_0) \right ]
\right \} \; ,
\ee
valid for $t_0\gg\tau_{loc}$, with a fitted relaxation rate $\Gm$.

There is no yet a quantitative theory describing nonequilibrium
condensate formation from the very beginning of the process, starting
from the kinetic stage, with no condensate, and going through the
heterophase and hydrodynamic stages.

\section{Low-Dimensional Condensates}

The properties of systems in one and two dimensions can essentially
differ from those in three dimensions. The Bose systems with the
contact interaction
\be
\label{36}
\Phi_d(\br) = A_d \dlt(\br) \; ,
\ee
where $\br$ is a $d$-dimensional vector, have been studied for
the spatial dimensionality $d=1$ and $d=2$ as well as for the usual
three-dimensional case. For the latter, one has $A_s=4\pi\hbar^2
a_s/m$.

First of all let us recall that the Bose-Einstein condensation in
uniform systems does not exist at finite temperatures in $d=1$ as
well as in $d=2$. This rigorously follows from the Bogolubov theorem
[72] as is shown in [102,103].

The features of low-dimensional Bose systems with the interaction (36)
can differ owing to the strength of their atomic interactions. This
can be illustrated by the classical example of the one-dimensional
Bose gas, solved by Lieb and Liniger [104,105]. Let us consider the
ground-state energy $E$, written in the dimensionless form as
$$
e(g) \equiv \frac{\hbar^2 n_1^2 E}{2m N} \qquad \left ( n_1 \equiv
\frac{N}{L} \right ) \; ,
$$ being a function of the dimensionless coupling
$$
g \equiv \frac{m A_1}{\hbar^2 n_1} \; .
$$
In the weak-coupling limit, when $g\ll 1$, the energy $e(g)$ possesses
an asymptotic expansion
$$
e(g) \simeq g + c_{3/2} g^{3/2} + c_2 g^2 + c_{5/2} g^{5/2} \; ,
$$
with the coefficients [106--108]
$$
c_{3/2}= -\; \frac{4}{3\pi}=-0.424413 \; , \qquad c_2=0.06535 \; ,
\qquad c_{5/2}=-0.017201 \; .
$$
In the strong coupling limit, with $g\ra\infty$, which is also called
the Tonks-Girardeau regime [109--111], one has
$$
e(\infty)= \frac{\pi^2}{3}=3.289868 \; .
$$
In the Tonks-Girardeau regime, the wave functions of the Bose system
are exactly mapped [110] by the Girardeau mapping $\Psi_B=|\Psi_F|$,
where $\Psi_F$ are the eigenstates of the ideal one-dimensional
Fermi gas provided by the simple Slater determinants. Thermodynamic
functions of the one-dimensional system are analytic at any finite
temperature [112], in agreement with the absence of a
finite-temperature phase transition [102,103].

Though there is no Bose-Einstein condensates in $d=1,2$ in
finite-temperature uniform systems, the properties of the latter
display noticeable change when varying temperature. This can be
demonstrated by considering the first-order density matrix
\be
\label{37}
\rho_1^{(d)}(\br) \equiv \; <\psi^\dgr(\br)\psi(0)> \; ,
\ee
in which the upper index $d$ shows dimensionality.

For the one-dimensional case, at zero temperature, one finds [113--115]
the large-distance decay
\be
\label{38}
\rho_1^{(1)}(\br) \sim r^{-\gm} \qquad (T=0) \; ,
\ee
where $\gm\equiv mc/2\pi\rho$ and $c=\sqrt{2A_d/m}$ is the sound
velocity. While at any finite temperature [115,116], for $r\ra\infty$,
one has
\be
\label{39}
\rho_1^{(1)}(\br) \sim \exp \left ( -\; \frac{m k_B T}{2n_1}\; r
\right ) \qquad (T>0) \; .
\ee
This means that there is no long-range order at any $T\geq 0$,
but at $T=0$ there arises {\it mid-range} order characterized by
the power-law decay (38). Therefore, at $T=0$, there can exist
superfluidity.

For the two-dimensional uniform gas at zero temperature the
long-range order does exist [5], since
\be
\label{40}
\rho_1^{(2)}(\br) \ra \rho_0 > 0 \qquad (T=0) \; ,
\ee
as $r\ra\infty$. At finite temperatures, below the
Kosterlitz-Thouless temperature $T_{KT}$, which is of the order of
the degeneracy temperature $T_d\sim\hbar^2\rho^{2/d}/m k_B$, the
matrix (37) decays algebraically [117--120],
\be
\label{41}
\rho_1^{(2)}(\br) \sim r^{-\gm} \qquad (0 < T < T_{KT} ) \; ,
\ee
similarly to the case (38), where $r\ra\infty$. Above the temperature
$T_{KT}$, the density matrix (37) decays exponentially, in analogy
with Eq. (39). Therefore, in the two-dimensional uniform Bose systems,
there is Bose-Einstein condensate at $T=0$, and there exists
superfluidity at $0<T<T_{KT}$, though there is no condensate at any
finite temperature.

The situation is different for low-dimensional {\it trapped} gases,
for which Bose condensation can occur at finite temperatures. Recall
that for a finite number of particles $N$ there is never a rigorous
phase transition, but Bose condensation happens as a gradual
crossover, which can be very sharp and practically undistinguishable
from the rigorous phase transition, when the number of particles $N$
is sufficiently large. This concerns trapped systems of any finite
dimensionality.

Phase transformations of the crossover type can occur, in some
cases, even  in the thermodynamic limit [95,121]. Therefore, it is
useful to generalize the notion of the phase transition, according
to the behaviour of the order parameter, as follows [95]. The phase
transition is of {\it first order}, when the order parameter changes,
at the transition point, by a jump. The phase transition is of {\it
second order}, when the order parameter at the transition point
changes continuously from zero to a finite value. And the phase
transition is of {\it third order}, or of {\it crossover type}, when
the order parameter changes continuously, being nonzero, so that there
occurs a qualitative change of its behaviour. The crossover point can
be defined as a point separating the qualitatively different regions.
Often, the crossover points can be identified with the locations of
the extrema of some derivatives [95,121,122].

Bose-Einstein condensation of trapped ideal gases in different
dimensions and in various traps has been reviewed in [2]. For $d=1$,
the quasiclassical approximation does not show any condensation
[2,123]. However, treating accurately the lower discrete levels
displays a crossover transition, which, for a harmonic trap of
frequency $\om_z$, can be located [124] at the temperature
\be
\label{42}
T_0^{(1)} = \frac{\hbar\om_z N}{k_B\ln(2N)} \; .
\ee

For a two-dimensional harmonic trap, with the radial frequency
$\om_\perp$, the quasiclassical approach is appropriate yielding
[2,123] the crossover temperature
\be
\label{43}
T_0^{(2)} = \frac{\sqrt{6N}\hbar\om_\perp}{\pi k_B} \; .
\ee
Below the temperatures (42) or (43) there occurs a macroscopic
occupation of the single-particle ground-state energy level, because
of which it is permissible to talk on Bose-Einstein condensation [2].
The crossover temperature $T_c$ in low-dimensional trapped dilute
gases with interaction is of the order of the temperatures (42) and
(43).

Dealing with interacting gases, it is not always easy to define the
single-particle spectrum, from which one would like to separate out
the ground-state level ascribed to Bose condensate [2]. Nevertheless,
in many cases, the condensate density can be defined, serving as
a reliable order parameter. A more general approach could be by
introducing the {\it order indices} [65,125]. An order index, for a
reduced $p$-particle density matrix $\hat\rho_p$, is defined [125] as
\be
\label{44}
\om_p \equiv \frac{\ln||\hat\rho_p||}{\ln|{\rm Tr}\hat\rho_p|} \; ,
\ee
where $||\cdot||$ implies the Hermitian norm. If the total number of
particles $N$ is large, one has $\ln|{\rm Tr}\hat\rho_p|\simeq p\ln N$.
For a pure Bose condensate, $\om_p=1$, while $\om_p=0$ when there is
no condensate. And in the intermediate region, the order index (44)
varies in the interval $0<\om_p<1$, being equal to $\om_p=\ln N_0/
\ln N$ [125].

Although in reality there are no exactly one-or two-dimensional
systems, it is possible, be varying the shape of the trapping
potentials, to produce dilute cold gases in highly anisotropic
configurations, where the motion of atoms is quenched in one or two
directions. The radial frequency $\om_\perp$ and the axial frequency
$\om_z$ of a cylindrical harmonic trap define, respectively, the
transverse and longitudinal oscillator lengths
$$
l_\perp \equiv \sqrt{\frac{\hbar}{m\om_\perp}} \; , \qquad
l_z \equiv \sqrt{\frac{\hbar}{m\om_z}} \; .
$$
When the trap is strongly squeezed in the radial direction, so that
$l_\perp\ll l_z$, one has the quasi-one-dimensional configuration.
And if the trap is squeezed in the axial direction, so that $l_z\ll
l_\perp$, one gets the quasi-two-dimensional system. These two opposite
configurations correspond to the cigar-shape or disc-shape traps. For
such highly anisotropic traps, it is possible to reduce the generic
three-dimensional consideration to an effective one- or two-dimensional
case, with the appropriate interaction constant $A_d$ in the potential
(36). If the scattering length $a_s$ is smaller, by modulus, than all
oscillator lengths, the reduction to a low-dimensional case is
straightforward, requiring just to integrate out the variables related
to the quenched directions [2]. Thus, for the quasi-one-dimensional
trap, one finds
\be
\label{45}
A_1 = \frac{2\hbar^2 a_s}{m l_\perp^2} \qquad
(|a_s|\ll l_\perp\ll l_z) \; .
\ee
And for the quasi-two-dimensional trap, one obtains
\be
\label{46}
A_2 = \sqrt{8\pi} \; \frac{\hbar^2 a_s}{ml_z} \qquad
(|a_s|\ll l_z \ll l_\perp ) \; .
\ee

The situation becomes more complicated when the scattering length is
comparable with an oscillator length. For example, if the scattering
length is of the order or even larger than $l_\perp$, then in the
quasi-one-dimensional case, the quartic self-interaction could be
replaced by a nonpolynomial interaction [126--128] or by a sextic term
[129,130]. A simple form of $A_1$ for arbitrary scattering lengths
was suggested by Olshanii [131],
\be
\label{47}
A_1 = \frac{2\hbar^2 a_s}{ml_\perp^2} \left [ 1 - \;
\frac{|\zeta(1/2)|a_s}{\sqrt{2}\; l_\perp} \right ]^{-1} \; ,
\ee
where $\zeta(1/2)\cong-1.46$. In the limit $|a_s|\ll l_\perp$, this
reduces to Eq. (45). Bose gases confined in highly elongated harmonic
traps were investigated over a wide range of interaction strengths
using quantum Monte Carlo techniques [132]. The properties of the Bose
gases under tight transverse confinement were found to be well
reproduced by a one-dimensional Hamiltonian with the contact interaction
(36) and the interaction parameter (47). The latter well represents
atomic interactions for arbitrary scattering lengths, positive as well
as negative. For $|a_s|\gg l_\perp$, the parameter (47) tends to the
unitary limit
\be
\label{48}
A_1 \simeq - 1.9368 l_\perp \hbar \om_\perp \qquad
(|a_s|\gg l_\perp) \; ,
\ee
which does not depend on $a_s$ and is attractive.

In the quasi-two-dimensional  case, the effective interaction parameter
for arbitrary $a_s$ was considered by Petrov et al. [133,134]. Their
results can be presented in the form
\be
\label{49}
A_2 =\sqrt{8\pi} \; \frac{\hbar^2 a_s}{m l_z} \left [ 1 +
\frac{2a_s}{\sqrt{2\pi}l_z} \; \ln\left (
\frac{l_\perp}{\sqrt{2\pi}l_z}\right ) \right ]^{-1} \; .
\ee
In the limit of the asymptotically small scattering length
$|a_s|\ll l_z$, Eq. (49) reduces to the form (46). However, under the
condition
$$
\frac{a_s}{l_z} \; \ln\frac{l_\perp}{l_z} \gg 1 \; ,
$$
the interaction parameter (49) becomes
\be
\label{50}
A_2 \simeq \frac{2\pi\hbar^2}{m\ln(l_\perp/\sqrt{2\pi}l_z)} \; ,
\ee
and does not depend on $a_s$. One could also introduce, instead of the
contact interaction (36), some more complicated interaction potentials
[130,135,136].

Note that for repulsive interactions, when $a_s>0$, the parameter (49)
is always positive. But for attractive interactions, when $a_s<0$, the
effective quasi-two-dimensional interaction (49) has a resonance form,
diverging at an intermediate scattering length
$$
a_s = -\; \frac{\sqrt{2\pi}\; l_z}{2\ln(l_\perp/\sqrt{2\pi}l_z)} \; ,
$$
and changing sigh from negative to positive.

Quasi-one-dimensional strongly alongated cigar-shaped traps have
been used in several experiments [137--145]. Quasi-two-dimensional
disc-shaped traps have also been employed in experiments [141,146].
For example, in experiments [142,143] the trap frequencies were
$\om_\perp=2\pi\times 365$ Hz and $\om_z=2\pi\times 14$ Hz. The
frequency ratio $\om_\perp/\om_z=26$ corresponds to the length ratio
$l_z/l_\perp=5$, which describes an alongated trap. The radial squeezing
can be increased [139], reaching $\om_\perp=2\pi\times 715$ Hz, which
translates into the length ratio $l_z/l_\perp=7$. Quasi-one-dimensional
condensates can also exist in waveguides [147].

\section{Crossover Region}

In the vicinity of critical points, there always exists a region,
where the system experiences strong fluctuations. This region is
termed critical. In finite systems, the phase transition is of
crossover type. Around the point of crossover, there can arise
strong fluctuations, similarly to those appearing in the vicinity
of the critical points. Then the region around a crossover point,
where fluctuations are strong, can be called crossover region.
Bose-Einstein condensation of trapped atoms is always a crossover.
In three dimensions, the latter can be rather sharp, when the number
of trapped particles is large. Therefore the crossover region can
be quite narrow. But in very anisotropic traps, in which the
quasi-one-dimensional or quasi-two-dimensional configuration is
realized, the condensation crossover can go as a very smeared
process, with a wide crossover region.

Considering phase transitions under varying temperature, one usually
defines the critical region by means of the Ginzburg temperature
$T_G$, such that in the critical region $T_G<T<T_c$ fluctuations
are strong, while essentially below $T_G$, they are suppressed [148].
In general, one may distinguish the longitudinal and transverse
fluctuations of an order parameter. The transverse, or directional,
fluctuations often survive much longer, when lowering temperature
below $T_c$, then the longitudinal, or size, fluctuations. Therefore
the actual critical region is characterized rather by the transverse
then longitudinal fluctuations [81].

In Bose-Einstein condensates, one may also distinguish two types
of fluctuations. The size fluctuations of the order parameter
correspond to density fluctuations. The transverse fluctuations
are usually associated with phase fluctuations. In the vicinity
of the condensation temperature $T_c$, both these fluctuations
are strong. Lowering temperature below $T_c$ first diminishes the
density fluctuations, while the phase fluctuations remain yet
strong. Lowering temperature further, one comes to a point $T_f$,
where phase fluctuations are getting suppressed as well. Thus, the
crossover region is defined by the temperature interval $T_f<T<T_c$.
For temperatures essentially lower than $T_c$, but yet much higher
than $T_f$, the system can be described as being spatially divided
into subvolumes, each of which being a coherent condensate, but
with no coherence between different subvolumes. Such a system is
an example of heterophase matter [95,149], in this particular case,
of a heterophase condensate. It has also been called quasicondensate
[150]. As is discussed in Section 4, the heterophase condensate,
or quasicondensate, appears in the process of nonequilibrium
condensation. And it may also exist in the crossover region
of equilibrium systems. The crossover regions are especially
pronounced in low-dimensional trapped gases [133,151,152],
though they also occur in purely three-dimensional cases [153].

To consider phase fluctuations, it is necessary, first of all, to
introduce a phase operator. In the works on Bose-Einstein condensation,
this is commonly done by presenting the field operator $\psi$ as
\be
\label{51}
\psi(\br) =\exp\left\{ i\hat\vp(\br)\right \}
\sqrt{\hat\rho(\br)} \;\; \; ?
\ee
Here $\hat\vp$ is the phase operator and $\hat\rho\equiv\psi^\dgr
\psi$. The question-mark in Eq. (51) is the sign of warning that
one should not blindly trust this equation, which, strictly speaking,
is not correct. The problem of introducing the phase operator is a
long-standing and not yet completely solved one, as can be inferred
from reviews [154--157] (see also a recent discussion in [158]). Thus,
in order to satisfy the identity $\hat\rho\equiv\psi^\dgr\psi$, the
phase operator is to be Hermitian, $\hat\vp^\dgr=\hat\vp$, so that
the creation field operator could be written as
\be
\label{52}
\psi^\dgr(\br) =\sqrt{\hat\rho(\br)} \exp\left \{ -i\hat\vp(\br)
\right \} \; \; \; ?
\ee
Here and in what follows, we continue marking by the question-mark
those equations that in strict sense are not correct. Above all, the
phase operator cannot be defined as a Hermitian operator [154--157].

The fact that the representation (51) is not correct can be
demonstrated as follows. From this representation, one has
$\exp(i\hat\vp)=\psi\hat\rho^{-1/2}$. Then, using the commutation
relation
$$
[\psi(\br),\; \hat\rho(\br')] = \psi(\br)\dlt(\br-\br') \; ,
$$
which is valid for any (either Bose or Fermi) field operators, one
gets
\be
\label{53}
[e^{i\hat\vp(\br)},\; \hat\rho(\br')] = e^{i\hat\vp(\br)}
\dlt(\br-\br') \; \; \; ?
\ee
From here, after expanding the exponentials, it follows the
commutation relation
\be
\label{54}
[\hat\rho(\br),\; \hat\vp(\br') ] = i\dlt(\br-\br') \; \; \; ?
\ee
This is what one standardly employs announcing the density and phase
operators being canonically conjugated. However, the relation (54) is
improper. Really, if it were correct, then for the number-of-particle
operator $\hat N\equiv \int\hat\rho(\br)d\br$ one would have
\be
\label{55}
[\hat N,\; \hat\vp(\br)] = i \;\;\; ?
\ee
The latter equation is already suspicious, with its left-hand side
depending on $\br$ and the right-hand side containing no such a dependence.
Moreover, Eq. (55) can be reduced to a completely ugly form by taking its
matrix element over the states $|n>$ from the number basis, for which
$\hat N|n>=n|n>$. This yields
\be
\label{56}
(n-n')\; <n|\hat\vp|n'>\; = i\dlt_{nn'} \; \; \; ?
\ee
Setting here $n=n'$ results in a senseless equation $0=i$.

In this way, the representation (51) cannot possess a rigorous
mathematical meaning. Then why people use it? To answer the question,
let us note that if, instead of the field operator $\psi$, we would deal
with a classical wave function, then the form $\psi=\sqrt{\rho}e^{i\vp}$
would be absolutely proper, but with the density $\rho$ and phase $\vp$
being also classical functions. This remark suggests that the
representation (51) could be accepted as an approximate form for
a kind of semiclassical approximation, when the density and phase
operators could be treated as almost classical functions. This is
equivalent to saying that the density and phase fluctuations above
their average values, for which $\dlt\hat\rho\equiv\hat\rho-<\hat\rho>$
and $\dlt\hat\vp\equiv\hat\vp-<\hat\vp>$, should be  very small. The
situation characterized by small fluctuations corresponds to the
hydrodynamic approximation.

There could be several equivalent ways of introducing a semiclassical
picture. One could treat the evolution equations for the mode operators
as semiclassical, in analogy with the semiclassical laser theory [159].
One could start from the Gross-Pitaevskii equation, present the wave
function in the form $\psi=\sqrt{\rho}e^{i\vp}$, and then quantize
the classical density and phase [4,160]. Or one could employ from the
very beginning the representation (51), but not forgetting that it is
only approximately meaningful in the hydrodynamic limit, when all
fluctuations are small. In this way, one often treats the density and
phase fluctuations as not correlated with each other, permitting their
separate averaging. And the phase operators are usually considered as
random Gaussian variables, which allows one to invoke, for a function
$f(\hat\vp)$ of $\hat\vp$, the equality
$$
<\exp\{ f(\hat\vp)\} > \; =
\exp\left\{ \frac{1}{2}\; < f^2(\hat\vp) > \right\} \; .
$$

It is worthwhile to emphasize that the phase, as such, is not an
observable quantity. What one is able to observe, except the density
of atoms, is the density of current, whose operator reads
$$
\hat{\bf j} \equiv -\; \frac{i\hbar}{2m} \left [
\psi^\dgr\nabla\psi - (\nabla \psi^\dgr)\psi \right ] \; ,
$$
or the velocity, for which the velocity operator is
$\hat{\bf v}=\hat{\bf j}/\rho$. For the classical quantities,
one would have $\hat{\bf j}=(\hbar\rho/m)\nabla\vp$ and
${\bf v}=(\hbar/m)\nabla\vp$. Respectively, instead of talking
about unobservable phase fluctuations, one could discuss the current
or velocity fluctuations, which can be correctly defined without
mentioning the problematic phase operator.

The density operator $\hat\rho$ and the operator of current
$\hat{\bf j}$ are connected by the continuity equation
$$
\frac{\prt\hat\rho}{\prt t} + {\rm div}\hat{\bf j} = 0 \; ,
$$
which is valid for an arbitrary trapping potential and any atomic
interactions. Hence, current fluctuations (or phase fluctuations)
would, generally, provoke density fluctuations, and vice versa, since
they are also connected by the continuity equation
$$
\frac{\prt}{\prt t} \dlt\hat\rho +
{\rm div}\dlt \hat{\bf j} = 0 \; .
$$
Density fluctuations always excite current fluctuations. However,
there can exist current fluctuations that do not initiate density
fluctuations. This takes place for the divergenceless fluctuations,
for which ${\rm div}\dlt\hat{\bf j}=0$. Hence, there can exist
circulating current fluctuations that leave the atomic density
undisturbed. This explains why the crossover region is mainly
defined by the phase (current) fluctuations, but not by the density
fluctuations.

In terms of the observable quantities, we can say that, in the
vicinity of the crossover temperature $T_c$, both the density and
current fluctuations are strong. In the part of the crossover region,
where $T_f<T\ll T_c$, the density fluctuations can be suppressed,
but the current fluctuations are yet essential, which is typical
of quasicondensate. And for $T\ll T_f$, true condensate develops.

For quasi-one-dimensional gases, the fluctuations temperature $T_f$
can be estimated [151,152] as
\be
\label{57}
T_f^{(1)} \approx \frac{\hbar\om_z}{k_B}\left (
\frac{l_\perp^2}{a_s l_z}\right )^{2/3} \; N^{1/3} \; .
\ee
The characteristic phase-fluctuation length is
\be
\label{58}
l_f^{(1)} \approx \frac{\hbar\om_z}{k_B T}\left (
\frac{l_\perp^2 l_z^2}{a_s}\right )^{1/3} \; N^{2/3} \; ,
\ee
which should be much less than the length $L$ of the trap.

In quasi-two-dimensional gases, the fluctuation temperature becomes
[133]
\be
\label{59}
T_f^{(2)} \approx \frac{\hbar\om_\perp}{k_B}\left (
\frac{l_z}{a_s}\right )^{1/2} \; \frac{\sqrt{N}}{\ln N} \; ,
\ee
with the fluctuation length
\be
\label{60}
l_f^{(2)} \approx \lbd_T \exp \left (
\lbd^2_T\rho^{2/3} \right ) \; ,
\ee
where it is assumed that $a_s\ll l_z$ and
$\lbd_T\equiv\sqrt{2\pi\hbar^2/m k_BT}$ is the thermal wavelength.
Here, it should be that $l_f^{(2)}\ll R$, with $R$ being the trap
radius.

The existence of fluctuations in the crossover region of elongated
Bose-Einstein condensates has been confirmed experimentally [142--145].

\section{Condensates and Superfluids}

The relation between Bose-Einstein condensation and superfluidity is
a long-standing and not completely understood problem. One often thinks
of these two phenomena as being closely related. A detailed description
of the historical evolution of views on this problem has recently been
done in [2]. A brief and nice discussion can also be found in [161].
The current understanding [2,161] is that Bose-Einstein condensation
is {\it neither necessary nor sufficient} for superfluidity. The latter
can occur without Bose-Einstein condensation. Thus, in two-dimensional
uniform systems, as is mentioned in Section 5, there can exist
superfluidity, though the condensate cannot appear at any finite
temperature. Huang [161] gives an example of a Bose system in a random
field, when superfluidity is destroyed, even though there is a Bose
condensate. Carusotto and Castin [162] find superfluid behaviour in a
one-dimensional ring of weakly interacting gas outside the Bose condensed
regime. In the present section, we shall adduce arguments why there can
be no simple and general relation between the densities of condensate and
of superfluid.

The condensate density
\be
\label{61}
\rho_0 =\rho -\rho_{out}
\ee
is defined as the difference between the total density and that of
atoms outside the condensate. The latter density for a uniform gas reads
\be
\label{62}
\rho_{out} =\frac{1}{V} \sum_{k\neq 0} n_k \; , \qquad n_k\equiv\;
<a_k^\dgr a_k>\; .
\ee
For instance, in the Bogolubov approximation at low temperature one finds
[70,71,163]
\be
\label{63}
\frac{\rho_0}{\rho} \simeq 1 -\; \frac{8}{3\sqrt{\pi}}\;
\sqrt{\rho a_s^3} - \; \frac{m(k_BT)^2}{12\hbar^3\rho c} \; ,
\ee
where $c=\sqrt{4\pi\hbar^2\rho a_s}/m$ is the Bogolubov sound velocity.

To find the superfluid density $\rho_s$, one consider a system boosted
with the velocity ${\bf v}=\{ v^\al\}$. In the frame, moving with the
same velocity, the Hamiltonian is a functional $H\{\psi\}$ of the field
operators $\psi$. In the laboratory frame, the Hamiltonian
$H_v=H\{\psi_v\}$ is the same functional, but of the field operators
$\psi_v$ given by the Galilean transformation
$$
\psi_v(\br,t) = \psi(\br,t) \exp\left\{ \frac{i}{\hbar}\left (
m{\bf v}\cdot\br -\; \frac{mv^2}{2}\; t \right ) \right \}.
$$
The superfluid component is interpreted as that part of the system,
which nontrivially responds to the velocity boost, with the superfluid
density
\be
\label{64}
\rho_s \equiv \frac{\rho}{mN}\; \lim_{v\ra 0}\; \frac{1}{3}\sum_\al
\frac{\prt}{\prt v^\al}\; <\hat P_v^\al>_v \; ,
\ee
where the averaging, with the Hamiltonian $H_v$, is over the total
momentum
$$
\hat{\bf P}_v \equiv \int \psi^\dgr_v(\br)(-i\hbar\nabla)
\psi_v(\br)\; d\br \; .
$$
Performing the limit $v\ra 0$ in Eq. (64), one has
\be
\label{65}
\rho_s = \rho -\rho_n \; ,
\ee
with the normal density
\be
\label{66}
\rho_n = \frac{\rho\bt}{3mN}\; \Dlt^2(\hat{\bf P}) \qquad
\left (\bt\equiv \frac{1}{k_B T} \right )
\ee
expressed through the momentum dispersion
$$
\Dlt^2(\hat{\bf P})\equiv <\hat{\bf P}^2>\; -\;
<\hat{\bf P}>^2 \; .
$$
When the total current across the system is absent, then
$<\hat{\bf P}>=0$. Combining Eqs. (61) and (65), one gets the formal
relation
\be
\label{67}
\rho_s = \rho_0 +\rho_{out} -\rho_n \; .
\ee

Note that the derivative in Eq. (64) is a particular case of
differentiation with respect to a parameter. In general, if a
parameter $\lbd$ enters an operator $\hat A=\hat A(\lbd)$ as well as
the Hamiltonian $H=H(\lbd)$, then
$$
\frac{\prt<\hat A>}{\prt\lbd} =\; <\frac{\prt\hat A}{\prt\lbd}> \; -
\bt \left ( <\hat A\; \frac{\prt H}{\prt\lbd}>\; - \; <\hat A>
<\frac{\prt H}{\prt\lbd}>\right ) \; .
$$
This formula of parametric differentiation is useful for different
particular applications.

To find the normal density (66) one needs to calculate the momentum
dispersion, which, for a motionless gas, is $\Dlt^2(\hat{\bf P})=
<\hat{\bf P}^2>$. This can be expressed [2] as
\be
\label{68}
<\hat{\bf P}^2>\; = \hbar^2 \int \lim_{r_3\ra r_1}
\lim_{r_4\ra r_2} \nabla_3 \cdot \nabla_4 \left [
\rho_1(\br_4,\br_3)\dlt(\br_1-\br_2) - \rho_2(\br_3,\br_4,\br_1,\br_2)
\right ]\; d\br_1 d\br_2 \; ,
\ee
where the first-and second-order density matrices, respectively, are
$$
\rho_1(\br_4,\br_3) =\; < \psi^\dgr(\br_3)\psi(\br_4)> \; ,
$$
$$
\rho_2(\br_3,\br_4,\br_1,\br_2) =\;
< \psi^\dgr(\br_2)\psi^\dgr(\br_1)\psi(\br_3)\psi(\br_4)> \; .
$$
Thus, the definition of the superfluid density (65) involves the
second derivatives of the first-and second-order density matrices,
while the condensate density (61) requires solely the first-order
density matrix. For interacting systems, there is no a simple general
relation between the reduced density matrices of different orders
(see [65] and recent discussions in [164--168]). This means that it
is hardly probable that there could exist a general relation between
the condensate density and superfluid density.

For Bose systems with condensate, one usually employs the
Bogolubov shift $\psi(\br)=\psi_0+\tilde\psi(\br)$, separating the
condensate part $\psi_0$ from the field operator $\tilde\psi(\br)$
of noncondensed particles. The condensate term $\psi_0$ is treated
as a classical function, which becomes a constant for uniform
gases. The field operator of noncondensed particles, for which
$<\tilde\psi>=0$, can be expanded over a basis $\vp_k(\br)$,
so that $\tilde\psi(\br)=\sum_{k\neq 0} a_k\vp_k(\br)$. For uniform
systems, which are considered in what follows, the basis is formed
by plane waves $\vp_k(\br)=e^{i{\bf k}\cdot\br}/\sqrt{V}$. In this
way, the single-particle density matrix reads
\be
\label{69}
\rho_1(\br_1,\br_2) = \rho_0 + \frac{1}{V} \sum_{k\neq 0}
n_k\; e^{i{\bf k}\cdot\br_{12}} \; ,
\ee
where $\br_{12}\equiv\br_1-\br_2$ and the property $<a_k^\dgr a_q>=
\dlt_{kq} n_k$ is taken into account.

Constructing the second-order density matrix in the Bogolubov approach,
one has to omit all terms containing the powers higher than those of
second order in the noncondensed field operators $\tilde\psi$ or $a_k$
with $k\neq 0$. Then, keeping in mind the momentum conservation, resulting
in the condition $<a_k a_q>=\dlt_{-kq}<a_k a_{-k}>$, one obtains
$$
\rho_2(\br_3,\br_4,\br_1,\br_2) = \rho_0^2 +
$$
\be
\label{70}
+ \frac{\rho_0}{V} \sum_{k\neq 0} \left \{ <a_k a_{-k} >\left (
e^{i{\bf k}\cdot\br_{12}} + e^{i{\bf k}\cdot\br_{34}}\right ) +
n_k \left ( e^{-i{\bf k}\cdot\br_{12}} + e^{-i{\bf k}\cdot\br_{14}}
+ e^{-i{\bf k}\cdot\br_{23}} + e^{-i{\bf k}\cdot\br_{24}}\right )
\right \} \; .
\ee
The diagonal element of the density matrix (70) gives the pair
correlation function
\be
\label{71}
g(\br_{12}) = \frac{1}{\rho^2}\; \rho_2(\br_1,\br_2,\br_1,\br_2) \; .
\ee
For the latter, when $N_0\approx N$, one finds
\be
\label{72}
g(\br) = 1 + \frac{2}{\rho} \int \left [ n_k +
<a_k a_{-k} > \right ]\; e^{i{\bf k}\cdot\br} \;
\frac{d{\bf k}}{(2\pi)^3} \; .
\ee
From here, one has the equality
$$
\int [ g(\br) - 1 ]\; d\br = \frac{2}{\rho} \; \lim_{k\ra 0} \left [
n_k + <a_k a_{-k}> \right ] \; ,
$$
which, in the Bogolubov approximation [70,71], yields
$$
\int [ g(\br) -1 ] \; d\br = \frac{1}{\rho} \left (
\frac{k_B T}{mc^2}\; - 1 \right ) \; ,
$$
with $c$ being the Bogolubov sound velocity. It is this equation that
was obtained by Bogolubov when calculating the particle dispersion
(22), with the use of Eq. (18). A similar way can be followed in the
calculation of the momentum dispersion (68), which is necessary for
finding the normal density (66).

Note that for the ideal gas it is easy to show that $\rho_0=\rho_s$
(see details in [2]). This result, however, is not of much meaning.
First, the ideal Bose gas cannot be superfluid, since it does not
satisfy the Landau criterion of superfluidity. Second, the ideal
Bose-condensed gas, as is explained in Sec. 3, is an unstable system,
having infinite compressibility.

It is also easy to show that, for any system at asymptotically high
temperatures, one has $\rho_n=\rho$, hence $\rho_0=\rho_s=0$. This
immediately follows from the Maxwell-Boltzmann distribution law
$n_k\simeq\exp\{\bt(\mu-\om_k)\}$, where, assuming the parabolic
spectrum $\om_k$, the chemical potential is
$$
\mu\simeq -\; \frac{3}{2}\; k_B T
\ln\; \frac{m k_B T}{2\pi\hbar^2\rho^{2/3}} \; .
$$
To illustrate the way of calculating the momentum dispersion (68)
for interacting gases, let us resort to the Bogolubov approach
[70--72]. Then, the annihilation and creation operators of noncondensed
particles in the momentum representation are $a_k$ and $a_k^\dgr$, with
${\bf k}\neq 0$. After omitting in the Hamiltonian all terms of third
and fourth order with respect to these operators, the remaining quadratic
form is diagonalized by means of the Bogolubov canonical transformation
$$
a_k = u_k b_k + v_k b^\dgr_{-k}\; , \qquad
a_k^\dgr = u_k b_k^\dgr + v_k b_{-k} \; ,
$$
where the operators $b_k$ correspond to excitations. As a result of
this diagonalization, one gets
$$
u_k =\frac{1}{\sqrt{1-B_k^2}} \; , \qquad
v_k = \frac{B_k}{\sqrt{1-B_k^2}} \; ,
$$
where
$$
B_k = \frac{\ep_k-\sqrt{\ep_k^2+m^2c^4}}{mc^2}
$$
and $\ep_k$ is the Bogolubov spectrum
$$
\ep_k =\left [ (c\hbar k)^2 +\left ( \frac{\hbar^2 k^2}{2m}\right )^2
\right ]^{1/2} \; ,
$$
which can also be presented in the form
$$
\ep_k = \left [ \left ( mc^2 +\frac{\hbar^2 k^2}{2m} \right )^2 -
m^2c^4 \right ]^{1/2} \; .
$$
For the coefficient functions $u_k$ and $v_k$, one has
$$
u_k^2 = \frac{\sqrt{\ep^2_k+m^2c^4}+\ep_k}{2\ep_k} \; , \qquad
v_k^2 = \frac{\sqrt{\ep^2_k+m^2c^4}-\ep_k}{2\ep_k} \; ,
$$
which yields the properties
$$
u_k^2- v_k^2 = 1 \; , \qquad u_k v_k = -\; \frac{mc^2}{2\ep_k} \; ,
\qquad u_k^2 + v_k^2 = \frac{\sqrt{\ep_k^2+m^2c^4}}{\ep_k} \; .
$$
For the momentum distribution of excitations, one finds
$$
\nu_k \equiv \; <b_k^\dgr b_k> \; =
\frac{1}{e^{\bt\ep_k} - 1} \; ,
$$
while for the momentum distribution of atoms, one obtains
$$
n_k \equiv \; <a_k^\dgr a_k>\; = (u_k^2 + v_k^2)\nu_k +v_k^2 \; .
$$
For the anomalous average $<a_k a_q>=\dlt_{-kq}<a_k a_{-k}>$, one gets
$$
<a_k a_{-k}> \; = u_k v_k (1 + 2\nu_k) \; .
$$
Note that the anomalous average is of the same order as $n_k$.

In the long-wave limit $k\ra 0$, one has $\ep_k\simeq c\hbar k$, so that
$$
\nu_k\simeq \frac{1}{\bt\ep_k} \; - \; \frac{1}{2} +
\frac{\bt\ep_k}{36} \; ,
$$
$$
n_k \simeq \frac{mc^2}{\bt\ep_k^2} + \frac{1}{2\bt mc^2} +
\frac{\bt mc^2}{36} \; - \; \frac{1}{2} \; , \qquad
<a_k a_{-k}> \simeq -\; \frac{mc^2}{\bt\ep_k^2} \; - \;
\frac{\bt mc^2}{36} \; .
$$
Then the sum of the normal and anomalous averages gives
$$
\lim_{k\ra 0}\left ( n_k +<a_k a_{-k}> \right ) =
\frac{1}{2}\left ( \frac{k_B T}{mc^2}\; - 1 \right ) \; .
$$
These properties are useful for calculating the momentum dispersion (68),
for which one finds
\be
\label{73}
<\hat{\bf P}^2> \; = \sum_k (\hbar k)^2 n_k -
N_0 \lim_{k\ra 0} (\hbar k)^2 <a_k a_{-k} > \; .
\ee
Introducing the notation for the mean kinetic energy
\be
\label{74}
E_{kin} \equiv \frac{1}{N} \sum_k \frac{(\hbar k)^2}{2m} \; n_k =
\frac{\hbar^2}{2m\rho} \int k^2 n_k\; \frac{d{\bf k}}{(2\pi)^3}
\ee
and taking account of the limit
$$
\lim_{k\ra 0} (\hbar k)^2 <a_k a_{-k}> = - mk_B T \; ,
$$
one comes to the expression of the normal density (66), which is
\be
\label{75}
\rho_n = \frac{2E_{kin}}{3k_B T} \; \rho + \frac{\rho_0}{3} \; .
\ee
Unfortunately, the mean kinetic energy (74) possesses an ultraviolet
divergence, since at $k\ra\infty$, one has $\ep_k\simeq(\hbar k)^2/2m$
and $n_k\simeq(mc/\hbar k)^4$. Consequently, the normal density and,
hence, the superfluid density (65) are not defined in the Bogolubov
approximation, while the condensate density (61) is perfectly defined.

The message of this section is that a condensate and a superfluid are
not synonyms but are rather different things that should not be confused
with each other.

\section{Collective Excitations}

Generally, collective excitations in dilute Bose gases can be of two
types: elementary excitations and topological modes. {\it Elementary
excitations} represent small oscillations above the ground state. They
are described by the linear Bogolubov-de Gennes equations. In uniform
gases, this yields the famous Bogolubov spectrum (see the previous Sec.
7). Elementary excitations leave the average spatial density unchanged.
Contrary to this, {\it topological modes} are strong excitations
essentially changing the spatial density distribution. These are
described by the nonlinear Gross-Pitaevskii equation. In uniform systems,
the known example of a topological mode is a vortex.

In nonuniform trapped gases, both these types of excitations are
also present, though the situation becomes much richer. Elementary
excitations in trapped gases can be subdivided in two kinds: continuous
modes and discrete modes.

When the excitation wavelengths are much shorter than the trap sizes,
the excitation spectrum is a continuous function of its wave vector.
This is why such excitations can be called {\it continuous modes}.
The measurement of the $k$ dependence of the excitation spectrum for
the trapped condensate of $^{87}$Rb atoms was accomplished in the nice
experiment by Steinhauer at al. [79]. The measured excitation spectrum
was found to be in very good agreement with the Bogolubov spectrum in
the local density approximation, even close to the long-wavelength
limit of the region of applicability. The static structure factor
$S(k)$ was also measured and found to agree well with the Bogolubov
theory.

A Monte Carlo computation of the static structure factor $S(k)$ of
a dilute Bose-Einstein condensate predicts a roton peak in $S(k)$
occurring at $k\approx 8\pi/a_s$, where $a_s$ is the $s$-wave
scattering length [169]. This peak is not steep enough to produce
a minimum in the excitation spectrum $\ep_k$.

For excitations with wavelengths $2\pi/k$ which are comparable
to the radius of the condensate, the excitation spectrum is
characterized by discrete shape-dependent oscillatory modes. Because
of this, such excitations can be termed {\it discrete modes}. For these
modes, momentum is not anymore a good quantum number and consequently
the elementary excitations are to be classified by the discrete quantum
numbers related to the trap symmetry [2,4]. The excitation spectrum of
discrete modes is mainly defined by the shape of the trapping potential,
essentially not depending on whether bosons or fermions are trapped
[170].

Discrete modes have been observed in several experiments.
Performing a small modulation of the trapping potential resulted
in the observation of center-of-mass oscillations [171--174]. Higher
multipolar-order surface oscillations were observed [175]. Scissors
modes were also studied in experiments [176,177]. More details can be
found in the review [2].

Theoretical calculations at zero temperature agree well with the
observed frequencies of discrete modes [2--4]. Temperature dependence
of the frequencies and of the corresponding attenuations, due to
interactions between the condensate collective modes and noncondensed
thermal atoms, have been theoretically analysed for uniform [178] and
trapped [179--182] gases. Finite-temperature calculations are in good
agreement with experiments for the lowest modes with axial angular
momentum quantum numbers $m=0$ and $m=2$ [183,184] and for scissors
modes [185,186].

The standard way of treating Bose-condensed systems is based on the
Bogolubov prescription $\psi=\psi_0+\tilde\psi$, with
$\psi_0\equiv<\psi>$, which breaks gauge symmetry. Then a Green
function approach is involved for either real-time or imaginary-time
Green functions, which are analytical continuations of each other on
the complex-time plane [187]. Nonequilibrium processes can be considered
by means of either Keldysh [91,188] or Kadanoff-Baym techniques [189,190].

Another approach is based on the Girardeau-Arnowitt canonical
transformation
$$
\al_k =\left ( \hat N_0 + 1 \right )^{-1/2}\; a_0^\dgr a_k \; ,
$$
in which $a_0$ is the condensate operator and $\hat N_0$ is the
condensate number-of-particle operator [191,192]. This approach was
followed in [181,193,194] and led to an accurate description of the
lowest discrete modes [184]. Note that Bogolubov [70,71] also used
a similar transformation $\al_k=N_0^{-1/2}a_0^\dgr a_k$, with the
difference that here $N_0$ is not an operator but the number of
condensed atoms. In both methods, a well developed condensate is
assumed to exist and in both the cases there appear the anomalous
averages $<a_k a_q>$. The resulting equations are also similar in
both these approaches.

A very popular trick is when, after breaking gauge symmetry
by means of the Bogolubov prescription, one omits all anomalous
averages. One usually ascribes this trick to Popov. However, this
is neither a correct approximation nor has any relation to Popov.
Really, as is easy to infer from the original works by Popov
[115,195,196], he never rejected the anomalous averages. Vise
versa, he emphasized the necessity of accurately taking them into
account in order to correctly calculate self-energies and to avoid
nonphysical singularities after breaking gauge symmetry by the
Bogolubov prescription.

Moreover it is straightforwarded to show that, generally, the anomalous
averages are of the same order as the normal ones, so that there is no
any reasonable justification of omitting the former though leaving the
latter [2,197,198]. For example, in the Bogolubov approach (see Sec. 7)
the normal average is
$$
<a_k^\dgr a_k> \; = \frac{\sqrt{\ep_k^2+m^2 c^4}}{\ep_k}\;
\left ( \frac{1}{2} + \nu_k \right ) - \; \frac{1}{2} \; ,
$$
where $\ep_k$ is the Bogolubov spectrum and
$\nu_k\equiv(e^{\bt\ep_k}-1)^{-1}$, while the anomalous average is
$$
<a_k a_{-k} > \; = -\; \frac{mc^2}{\ep_k} \; \left ( \frac{1}{2} +
\nu_k \right ) \; .
$$
These averages are closely related usually being of the same order.
Thus, in the long-wave limit $k\ra 0$, one has $<a_k^\dgr a_k>\simeq
mc^2/\bt\ep_k^2$ and $<a_k a_{-k}>\simeq-mc^2/\bt\ep_k^2$. That is,
in this case, the absolute values of the normal and anomalous averages
coincide at all temperatures below $T_c$. As soon as the anomalous
averages appear, they must be considered together with the normal
averages. There is no any justification of blindly throwing them out.

Then one may ask: if the trick of throwing out anomalous averages is
wrong, why, being used by so much people, it often yields reasonable
results? For instance, it leads to accurate calculated frequencies of
lower collective excitations in the temperature interval $0<T<0.6T_c$,
though it is not correct for the $m=2$ mode above $0.6T_c$. The answer
to this question is as follows.

For a system with Bose-Einstein condensate, it is possible to employ
a representation, based on {\it mixed coherent states} [60,199], where
the gauge symmetry is not broken and, hence, the anomalous averages
do not appear at all [125]. In this approach the field operator is
presented as
\be
\label{76}
\psi(\br,t) = \eta(\br,t) e^{i\al} +\tilde\psi(\br,t) \; ,
\ee
where $\eta(\br,t)$ is a coherent field and $\al\in[0,2\pi]$ is a random
phase. The average of an operator $\hat A$ is defined by
\be
\label{77}
<\hat A> \; \equiv \int_0^{2\pi} {\rm Tr}\hat\rho\hat A\;
\frac{d\al}{2\pi} \; ,
\ee
with $\hat\rho$ being a statistical operator. Because of the
averaging over the random phase, all anomalous averages are identically
zero, so that $<\psi>=<\tilde\psi>=0$ as well as $<\psi\psi>=<\tilde\psi
\tilde\psi>=0$ and so on. The equation for the coherent field $\eta=
\eta(\br,t)$ reads
\be
\label{78}
i\hbar\; \frac{\prt\eta}{\prt t} = \left [ -\;
\frac{\hbar^2}{2m}\; \nabla^2 + U + A_s \left ( |\eta|^2 +
2\tilde\rho \right ) - \mu \right ] \; \eta \; ,
\ee
where $U=U(\br,t)$ is a trapping (external) potential,
$A_s\equiv 4\pi\hbar^2 a_s/m$, and $\tilde\rho=\tilde\rho(\br,t)$ is
the density of noncondensed atoms. Equation (78) reminds a generalized
Gross-Pitaevskii equation, but it should be stressed that Eq. (78)
is an {\it exact} equation, with no approximation yet being involved.
The density of noncondensed atoms can be found from the related
single-particle Green function
$$
\tilde G(12) \equiv - i
<\hat T\tilde\psi(1)\tilde\psi^\dgr(2) > \; ,
$$
in which $\hat T$ is the chronological operator and each number
$j$ implies a set $\br_j,t_j$. The evolution equation for $\tilde G$
includes the corresponding two-particle Green function $\tilde G_2$
and self-energy
$$
\tilde\Sigma(12) = A_s \left [ 2\dlt(12) |\eta(1)|^2 +
i \int \tilde G_2(1113) \tilde G^{-1}(32)\; d(3) \right ] \; .
$$
This allows us to write the equation for $\tilde G$ as
\be
\label{79}
\left ( i\hbar \; \frac{\prt}{\prt t_1} + \frac{\hbar^2}{2m} \;
\nabla_1^2 - U + \mu \right ) \; \tilde G(12) - \int
\tilde\Sigma(13) \tilde G(32)\; d(3) = \dlt (12) \; ,
\ee
which is also an exact equation. These equations (78) and (79), being
exact, at the same time contain no anomalous averages. More details
can be found in Ref. [125]. Linearizing the coherent-field equation
(78) with respect to the deviation $\dlt\eta\equiv\eta-\eta_0$, where
$\eta_0$ corresponds to the ground state, it is possible to come to
the Bogolubov-de Gennes type equations describing elementary collective
excitations.

In the stationary situation, when the external field $U=U(\br)$ does not
depend on time, the coherent field can be presented as
$$
\eta(\br,t) =\sqrt{N_0}\; \vp(\br) \; \exp\left\{ -\;
\frac{i}{\hbar}\; (E + \mu) t\right \} \; .
$$
Then, Eq. (78) reduces to the stationary eigenvalue problem
\be
\label{80}
\hat H[\vp]\vp(\br) = E\vp(\br) \; ,
\ee
with the effective nonlinear Hamiltonian
$$
\hat H[\vp] \equiv -\; \frac{\hbar^2}{2m}\; \nabla^2 + U + A_s
\left ( N_0 |\vp|^2 + 2\tilde\rho \right ) \; .
$$
The eigenfunctions $\vp_n$ of Eq. (80) are termed {\it topological
coherent modes} [200--203]. The mode, with the lowest energy $E_0$,
describes the standard Bose-Einstein condensate, while the modes,
with higher energies $E_n>E_0$, correspond to excited nonground-state
condensates [200--203]. The generation of topological coherent modes
can be realized by applying alternating fields with frequencies in
resonance with the desired transition frequencies
$\om_{mn}\equiv(E_m-E_n)/\hbar$. The most known example of a topological
mode is a vortex, which can also be excited by means of resonant external
fields [204]. Reviews on vortices in trapped condensates can be found in
[205,206]. A dipole topological mode was generated in experiment [207].
The properties of different topological modes were studied in several
theoretical papers [200--203,207--221]. Such modes can also be generated
by a resonant modulation of the scattering length [222--224]. The
topological modes can be excited in single-well potentials, in optical
lattices [225], and in other multiwell potentials [226,227]. Adiabatic
preparation and evolution of two tunneling modes was investigated
[228--230]. Thermodynamics of an ensemble of the coherent modes were
described [231].

To stress more the principal difference between elementary excitations
and topological modes, let us consider the single-particle density
matrix $\hat\rho_1$, which can be presented as an expansion over its
eigenfunctions $|n>$, called natural orbitals [65], so that
$\hat\rho_1=\sum_n N_n|n><n|$, where the eigenvalue $N_n\equiv<a_n^\dgr
a_n>$ is the occupation number. In the case of a system with the usual
Bose-Einstein condensate, the ground-state occupation number $N_0\sim N$
becomes macroscopic, while all other levels, corresponding to elementary
excitations, are not macroscopically occupied, so that $N_n/N\ra 0$,
as $N\ra\infty$ and $n=0$. Contrary to this, for a system with several
topological modes, several occupation numbers $N_n$ become macroscopic.
This type of condensate is, sometimes, called fragmented [232]. Such
a fragmentation in the spectrum of the single-particle density matrix
should not be confused with a spatial fragmentation of a complex system
[233]. Thus, a system with several topological modes is an example of
the fragmented condensate in the sense of the spectral fragmentation
[232]. It is also worth noting that the multimode fragmented condensate
is, generally, nonequilibrium, since its realization requires an action
of alternating external fields or a special initial preparation.

\section{Particle Correlations}

Trapped atoms usually form a very dilute gas, such that $\rho a_s^3
\sim 10^{-8}-10^{-3}$. The quantity $\rho a_s^3$ can be increased either
by increasing the density $\rho$ or the scattering length $a_s$. The latter
can be varied by manipulating external magnetic fields close to Feshbach
resonance [234]. These are collision resonances that occur when the energy
of a colliding channel coincides with the energy of a long-lived bound state.
When the colliding atoms and the bound state have different magnetic momenta,
the resonance condition may be tuned via external magnetic fields exploiting
the Zeeman effect [2,235]. If the parameter $\rho a_s^3$ is essentially
increased, then interparticle correlations could become important and
even the whole picture based on the local interaction, involving just a
scattering length, could turn invalid. Correlations become crucial, making
the simple mean-field approximation unreliable, already at $\rho a_s
\sim 10^{-2}$ [54,236].

Feshbach resonances, first, were observed in Bose-Einstein condensates
of $^{85}$Rb, near the magnetic field $B_0=160$ G with width $\Dlt B=-6$
G [237], and of $^{23}$Na, near $B_0=907$ G with $\Dlt B=1$ G [238].
Then Feshbach resonances were realized in other bosonic atomic species
[239--242], as well as in several fermionic atoms [243--246]. The
behaviour of the scattering length near a Feshbach resonance at magnetic
field $B_0$ is given by the formula
\be
\label{81}
a_s(B) = a_s \left ( 1 -\; \frac{\Dlt B}{B-B_0} \right ) \; ,
\ee
in which $a_s \equiv a_s(\infty)$ is the off-resonance scattering length
and $\Dlt B$ characterizes the width of the resonance. Typical behaviour
of the relative scattering length $a_s(B)/a_s$ is illustrated in Fig.
1 for $^{85}$Rb, with $B_0=156$ G and $\Dlt B=-6$ G, and in Fig. 2 for
$^{23}$ Na, with $B_0=907$ G and $\Dlt B=1$ G. These figures follow
directly from Eq. (81). In the case of $^{85}$Rb, it was experimentally
possible to change $a_s(B)$ by about three orders of magnitude [237].
But for $^{23}$Na, the variation of $a_s(B)$ was limited by one order
because of the strong enhancement of inelastic collisions close to the
Feshbach resonance leading to an essential depletion of the condensate
[238,247]. As is seen in Figures 1 and 2, it is possible not merely
vary the scattering-length magnitude but even to change its sign. Such
experiments were carried out with $^{85}$Rb, when the sign of $a_s(B)$
was fastly changed from positive to negative, which led to the explosion
of the condensate [248,249].

When atomic interactions are so strong or density is so large that
$\rho a_s^3>0.1$, then it becomes crucial to take account of atomic
correlations which are well characterized by the pair correlation
function $g(\br,\br')$ defined in Sec. 3. In many cases, it is sufficient
to consider the uniform approximation $g(\br)$ depending on the difference
of spatial variables. It may also happen that the contact potential is
no longer representing interactions correctly and, instead of the contact
potential proportional to $\dlt(\br)$, one should consider a more realistic
form $\Phi(\br)$.

From the quantum scattering theory [250] it is known that, generally,
the scattering amplitude is expressed through the integrals of the type
$\int\vp_m^*(\br)\Phi(\br)\vp_n(\br)d\br$, where $\vp(\br)$ is a function
characterizing the relative motion of two colliding particles and
satisfying the Schr\"odinger equation
\be
\label{82}
\left [ - \; \frac{\hbar^2}{2m_{eff}} \; \nabla^2 +
\Phi(\br) - E \right ] \; \vp(\br) = 0 \; ,
\ee
in which $m_{eff}\equiv m_1m_2/(m_1+m_2)$ is the reduced mass of the
colliding particles with masses $m_1$ and $m_2$. For identical particles,
with $m_1=m_2\equiv m$, the reduced mass $m_{eff}=m/2$. It is only for
the low-energy Born approximation when the scattering amplitude contains
just the integral $\int\Phi(\br)d\br$, which reduces to a scattering
length.

The simplest way of defining the pair correlation function would be by
setting $g(\br)\equiv|\vp(\br)|^2$, with $\vp(\br)$ defined by Eq. (82),
but with the boundary condition $|\vp(\br)|\ra 1$ as
$r\equiv|\br|\ra\infty$, instead of the standard normalization condition.
Such a definition of $g(\br)$ was suggested by Bogolubov [70] and employed
for describing quantum crystals [251--257], quantum liquids [258,259],
melting-crystallization phase transformation [260], and, generally, for
considering quantum statistical systems with strongly singular interaction
potentials [261]. The parameter $E$ in Eq. (82) has to be chosen from some
additional physical conditions. Thus, it is natural to assume that the
pair correlation function $g(\br)=|\vp(\br)|^2$ possesses a maximum at
$\br_0$ where the interaction potential $\Phi(\br)$ has a minimum. Then
one should set $E=\Phi(\br_0)$. For a short-range repulsive potential,
one may set $E=0$.

For dense systems, it is possible to develop more refined ways of
constructing the pair correlation function $g(\br)$. It is also possible
to find $g(\br)$ from experiments [62,262] by measuring the structural
factor.

Including interparticle correlations from the very beginning implies
that the interaction potential $\Phi(\br)$ should enter everywhere
together with the pair correlation function $g(\br)$, forming the
smoothed potential $\overline\Phi(\br)\equiv\Phi(\br)g(\br)$, which
plays the role of a pseudopotential. The latter is always finite and
integrable, since, when $\Phi(\br)$ becomes singular, the correlation
function $g(\br)$ tends to zero, so that the smoothed potential
is always nonsingular, which removes the ultraviolet divergencies
appearing in the standard consideration starting with a simple
mean-field approximation, such as the Hartree-Fock approximation. An
approach, involving at all iteration steps only the smoothed potential
$\overline\Phi(\br)$, was, first, suggested in [263], which has been
developed into a refined {\it correlated iteration theory}
[60,264,265]. Here we present only the final set of equations of this
theory.

For the single-particle Green function
$$
G(12) \equiv - i\; <\hat T\psi(1)\psi(2)> \; ,
$$
we may write the Dyson equation
\be
\label{83}
G(12) = G_0(12) + \int G_0(13) \left [ \Sigma(32) -
\Sigma_0(32) \right ] \; d(3) \; ,
\ee
in which $G_0(12)$ is any trial Green function associated with the
self-energy $\Sigma_0(12)$. Here the same abbreviated notation as in
Sec. 8 is used. The self-energy
\be
\label{84}
\Sigma(12) = \pm i \int \overline\Phi(13) D(132)\; d(3)
\ee
is expressed through the smoothed potential
\be
\label{85}
\overline\Phi(12) \equiv g(12)\Phi(12) \; ,
\ee
where $g(12)$ is the pair correlation function, and the doubling
function $D(123)$ is given by the form
\be
\label{86}
D(123) = \hat Y D_0(123) \; ,
\ee
where the starting doubling function is
$$
D_0(123) \equiv \dlt(13) G(22) \pm G(12)\dlt(23) \; ,
$$
and the operator $\hat Y$ is defined below. For generality, Bose as well
as Fermi statistics are implied here, with the upper sign being related
to Bose, while the lower sign, to Fermi statistics.

An iterative solution of Eqs. (84) and (86) follows the scheme
\be
\label{87}
D_k \ra \Sigma_{k+1} \ra \hat Y_{k+1} \ra D_{k+1} \; ,
\ee
where the $k$-order approximation for the operator $\hat Y$ is
\be
\label{88}
\hat Y_k \equiv 1 + \sum_{p=1}^k \hat y_k^p \; ,
\ee
and the action of the operator $\hat y_k$ on any three-point function
$f(123)$ is defined as
$$
\hat y_k f(123) \equiv \left [ 1 - g(12) \right ] f(123) +
$$
\be
\label{89}
+ \int G(14)\; \frac{\dlt\Sigma_k(43)}{\dlt G(56)} \; \left [
g(52) f(527) G(76) - G(22) G(56)\dlt(67)\right ] \; d(4567) \; .
\ee
At any approximation order, particle interactions enter always in the
form of the smoothed potential (85) renormalized by the pair correlation
function $g(12)$. Thus, particle correlations are taken into account
from the very first step of the correlated iteration theory [60,264,265].

An analogous procedure can be followed for finding the response functions
[60]. For instance, let us consider the response function
\be
\label{90}
\chi(123) \equiv \mp i\; \frac{\dlt G(12)}{\dlt\mu(3)} \; ,
\ee
where $\dlt\mu(1)$ is a variation of the chemical potential. The poles
in the energy representation of the function $\chi(12)\equiv\chi(112)$
describe the spectrum of collective excitations. The response function
(90) can be presented as the action of an operator $\hat Z$ on the
initial function
$$
\chi_0(123) \equiv \pm i G(13)G(23) \; ,
$$
corresponding to the random-phase approximation, so that
\be
\label{91}
\chi(123) = \hat Z \chi_0(123) \; .
\ee
An iterative solution of the latter equation is realized according to
the scheme
\be
\label{92}
\Sigma_k \ra \hat Z_k \ra \chi_k \; ,
\ee
in which the $k$-order approximation of the operator $\hat Z$ is
\be
\label{93}
\hat Z_k \equiv 1 + \sum_{p=1}^k \hat z_k^p \; ,
\ee
with the action of $\hat z_k$ prescribed by the rule
\be
\label{94}
\hat z_k f(123) \equiv \int G(14) G(52) \;
\frac{\dlt\Sigma_k(45)}{\dlt G(67)}\; f(673)\; d(4567) \; .
\ee
Here again, we never meet the bare interaction potential, since the
self-energy $\Sigma_k(12)$ contains always only the smoothed potential
(85).

Wishing to take an explicit account of the arising Bose-Einstein
condensate, we may proceed as in Sec. 8, shifting the total field
operator as in Eq. (76) and separating an effective wave equation for
the coherent field, as in Eq. (78), from the equation for the Green
functions of noncondensed particles. However, a more general way is to
present the total Green function as the sum
\be
\label{95}
G(12) = - i\eta(1)\eta^*(2) + \tilde G(12) \; .
\ee
Note that the shift (76) is sufficient, though not necessary, for the
representation (95). A system, characterized by the Green function (95),
contains a coherent part, related to the coherent field $\eta$, and a
noncoherent part of noncondensed particles corresponding to $\tilde G$.
The existence of a coherent part can be detected experimentally [266].

The full many-body theory, accurately taking into account particle
interactions and correlations, will become necessary as soon as the
parameter $\rho a_s^3$ is not anymore small. As is explained at the
beginning of this section, this parameter can be made sufficiently
large, up to $\rho a_s^3\sim 1$, by tuning atomic interactions with
an external magnetic field close to a Feshbach resonance. In the
vicinity of the latter, there exist bound states, when atoms can
form molecules, whose interactions are long-ranged, also requiring
to accurately take into account particle correlations [267].

\section{Molecular Condensation}

Since there exist bound or quasibound states for colliding cold atoms,
it has been an important challenge to produce ultracold molecules that
could form Bose-Einstein condensates. Such molecules would behave as
bosons, independently of whether they are composed of bosons or fermions.

One route for producing ultracold molecules is to form them from Bose
atoms in the condensed state. Thus, a two-photon stimulated Raman
transition in the condensed $^{87}$Rb was used to produce $^{87}$Rb$_2$
molecules in a rotational-vibrational state [268], and ultracold molecules
of $^{23}$Na$_2$ were formed [269] through photoassociation of the
condensed $^{23}$Na. The molecules of $^{23}$Na$_2$ were also produced
by means of the Feshbach resonance technique [270], which was used as
well for producing the molecules $^{133}$Cs$_2$ [271]. Employing the
Feshbach resonance was found to be the most convenient way of creating
cold molecules, and this technique was used in all experimental works
cited below. Molecules are formed by sweeping an external magnetic
field through the Feshbach resonance, adiabatically converting atoms
to molecules.

The detection of whether the created molecules form a Bose-Einstein
condensate is a demanding problem requiring to probe coherence
properties of the assembly of atoms plus molecules. It was found
that the molecules $^{85}$Rb$_2$ do form the condensate [16] as well
as the molecules $^{87}$Rb$_2$ [272]. The molecules made of cold Bose
atoms are in highly vibrationally excited states and usually undergo
fast decay during $10^{-4}$ s.

Molecules can also be composed of cold Fermi atoms by tuning the
atomic interactions with a varying magnetic field close to a Feshbach
resonance. It is possible to vary the $s$-wave scattering length $a_s$
from positive to negative over many orders of magnitude. This opens
up the possibility of investigating the crossover from a molecular
condensate to atomic superconductivity [273,274]. The condensate limit
occurs for a positive scattering length $a_s>0$, much smaller than the
interatomic spacing, so that $\rho a_s^3\ll 1$. Then the fermions form
weakly bound dimers of a size about $a_s$. The superconducting limit
is expected to occur for a negative scattering length $a_s<0$, when
the distance between fermions in a superconducting pair is much larger
than the interatomic spacing. Then the Fermi momentum $k_F$ is such that
$k_F|a_s|\ll 1$. The Fermi temperature of a harmonically trapped gas
is $T_F\sim(\hbar\om_0/k_B)(3N)^{1/3}$. The superconducting transition
is expected to happen when $T/T_F\sim 10^{-2}-10^{-1}$. Thus, if $T_F
\sim 10^{-6}$ K, one needs to reach the temperatures $T\sim 10^{-7}$
K, which is achievable in modern experiments.

The molecules $^6$Li$_2$, composed of cold Fermi atoms $^6$Li, where
formed near a Feshbach resonance [275--277], being rather long-lived,
with a lifetime around 1 s. The molecules $^{40}$K$_2$ live shorter,
about $10^{-3}$ s [278]. It has been possible to convert to molecules
up to 80$\%$ of atoms.

Molecules, formed from Fermi atoms, behave themselves as bosons and,
hence, can be condensed at sufficiently low temperatures. Thus, the
molecules $^6$Li$_2$ condense at $T_c=6\times 10^{-7}$ K; below this
temperature condensate fraction of up to $90\%$ has been observed
[17,279--282]. The molecules $^{40}$K$_2$ can also be condensed, though
their condensate fraction being about $10\%$ [283,284]. It is especially
interesting to study the crossover region around a Feshbach resonance,
going from positive scattering lengths to negative ones. Then the
molecular condensate transforms into a degenerate Fermi gas of atoms.
This crossover region was investigated for $^6$Li$_2$ [280--282] as
well as for $^{40}$K$_2$ [284]. In the latter paper, condensed molecules
of $^{40}$ K$_2$ were observed in the region where the scattering length
is negative and no stable $^{40}$K$_2$ molecules would exist in vacuum.
It was claimed [284] that those condensed pairs were not bound into
molecules, but merely moved together in a correlated fashion, similar
to Cooper pairs of electrons in a superconductor, hence the claim was
that the superconductivity of $^{40}$K atoms had been observed, although
the exact nature of these pairs remained unclear. However, theoretical
calculations showed [285] that these pairs are not the Cooper pairs but
are the same tightly-bound molecules as those existing in the condensate
region, where $a_s>0$. The existence of such molecules in the degenerate
Fermi-gas region, where $a_s<0$, is due to the interaction of these
molecules with the unbound atoms of the degenerate Fermi gas. This
conclusion was confirmed by the experiment [282] with condensed
molecules $^6$Li$_2$, where the large condensate fraction, up to $80\%$,
was observed in the degenerate-gas region, where no stable $^6$Li$_2$
molecules could exist in vacuum. The existence of these molecules has
been interpreted as been stabilized by the presence of the Fermi gas
of unbound atoms [282].

The situation, when bound states, or bound clusters, that cannot exist
in vacuum, nevertheless, can perfectly survive inside a sea of their
unbound constituents is not unique for trapped atoms [122]. The same
picture happens for hadrons that are stable below the point of
deconfimenent but which would not be stable above this point, if they
would be placed in vacuum. Nonetheless, a fraction of hadrons can
perfectly survive above the deconfinement point provided they are
immersed into quark-gluon plasma [121, 122].

Concluding this review, I would like to repeat what was said in
the Introduction. The physics of trapped cold atoms is a very wide and
fast developing field. A number of interesting and important subjects
are not touched here, such as spinor condensates, dipolar condensates,
multicomponent condensates, and condensates in optical lattices. I have
concentrated in the present review on the most fundamental, to my
understanding, problems. And I hope to cover other topics, not touched
here, in another publication.

\vskip 5mm

{\bf Acknowledgement}

\vskip 2mm

I am grateful for many useful discussions and collaboration on some
topics of Bose-Einstein condensation to V.S. Bagnato, Ph.W. Courteille,
K.-P. Marzlin, and E.P. Yukalova. Financial support from the
Heisenberg-Landau Program (Germany) and from the German Academic Exchange
Service is appreciated.

\newpage

\begin{center}
{\large{\bf Figure Captions}}
\end{center}

\vskip 2cm

{\bf Fig. 1}.  Relative scattering length $a_s(B)/a_s$ for $^{85}$Rb close
to the Feshbach resonance at $B_0=156$ G with $\Dlt B=-6$ G.

\vskip 1cm

{\bf Fig. 2}. Relative scattering length $a_s(B)/a_s$ for $^{23}$Na close
to the Feshbach resonance at $B_0=907$ G with $\Dlt B=1$ G.

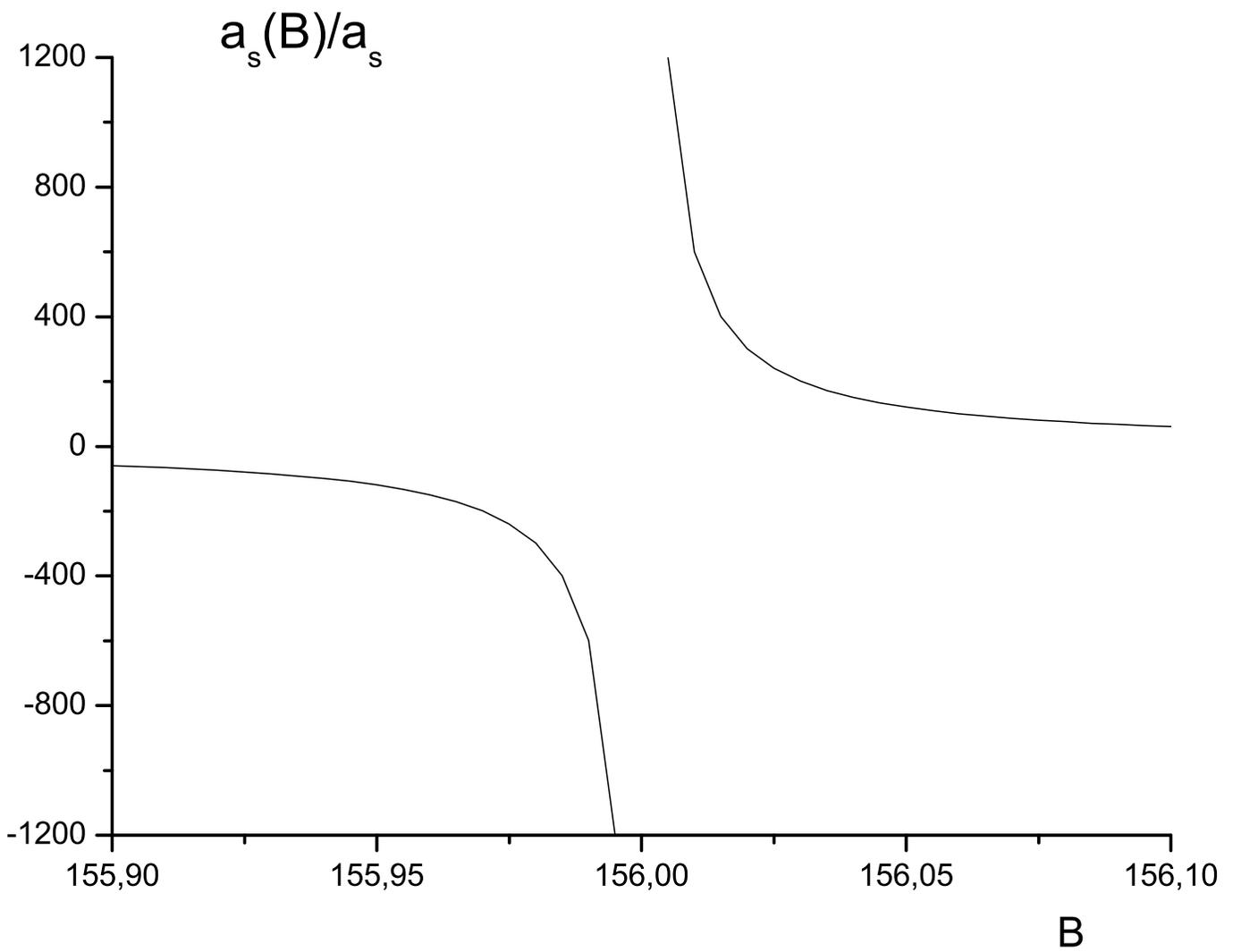
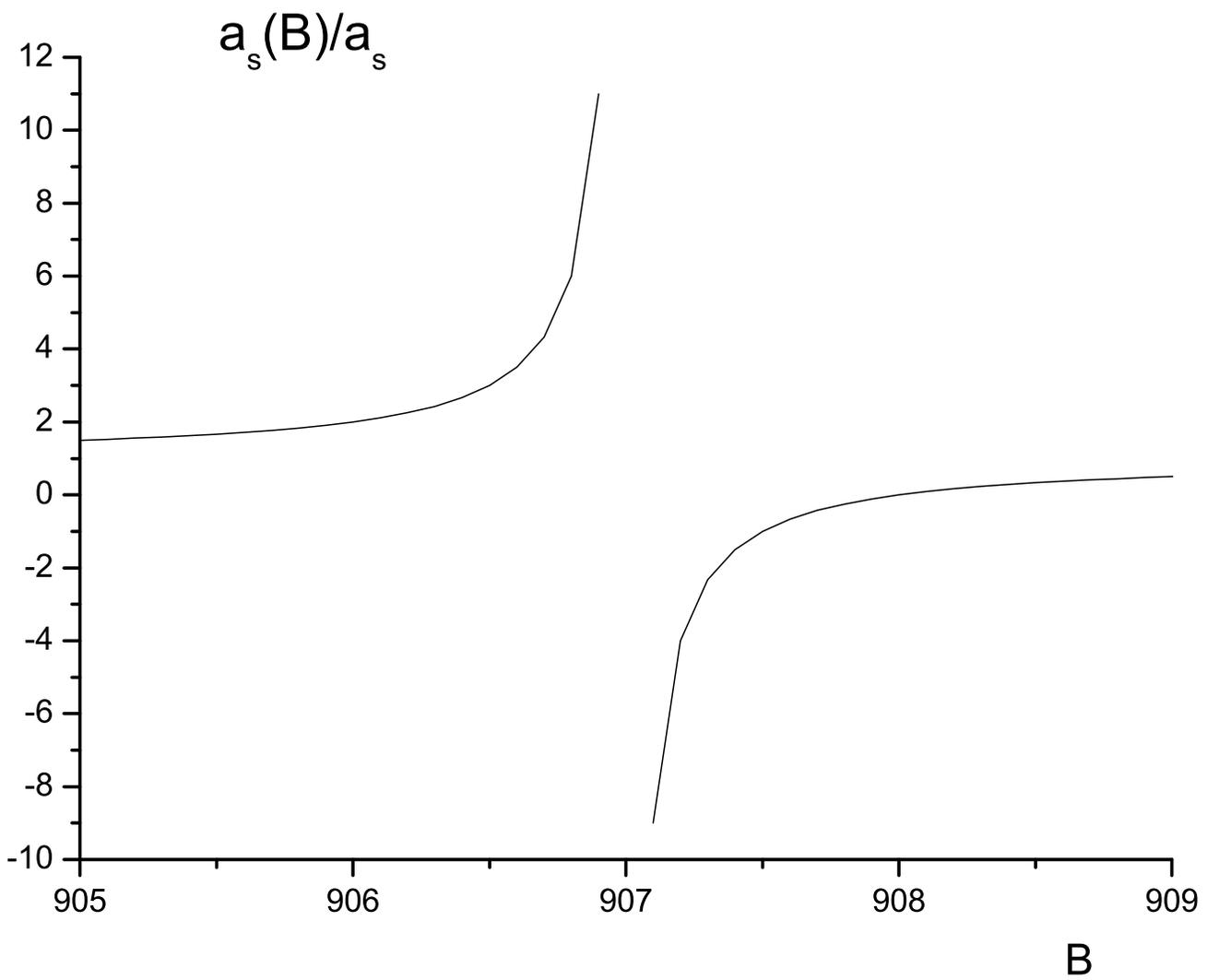
\end{document}